\documentclass[a4paper,10pt]{article}

\pdfoutput=1 

\usepackage{jheppub} 

\usepackage[utf8]{inputenc}
\usepackage[T1]{fontenc}
\usepackage[english]{babel}
\usepackage{fancyhdr}
\usepackage{booktabs}
\usepackage{hyperref} 
\usepackage{amssymb,amsfonts,amsmath}
\usepackage{graphics,graphicx,pstricks,color,colortbl}
\usepackage{slashed}
\usepackage{cleveref}
\crefname{figure}{Figure}{Figures}
\usepackage{multirow}
\usepackage{booktabs}
\usepackage{axodraw2}
\usepackage{braket}
\usepackage{amsmath}
\usepackage{comment}
\usepackage{graphicx}
\usepackage{tikz}
\usepackage{xcolor}
\usepackage{float}
\usepackage[T1]{fontenc}
\usepackage{placeins}
\usepackage{amsfonts}
\usepackage{mathrsfs}
\usepackage{bm}
\usepackage{subfig}
\usepackage[normalem]{ulem}

\newcolumntype{P}[1]{>{\centering\arraybackslash}p{#1}}

\newcommand{\GeV}{{\rm\ GeV}}

\newcommand{\snr}{{$\rm{SNR}>10$}}

\newcommand{\bl}[1]{\textcolor{black}{#1}}

\allowdisplaybreaks

\newcommand{\bpone}{\tikz[baseline=-0.6ex]\filldraw[fill=yellow,draw=black] (0,0) circle (0.9ex);}
\newcommand{\bptwo}{\tikz[baseline=-0.6ex]\filldraw[fill=green,draw=black] (-0.9ex,-0.9ex) rectangle (0.9ex,0.9ex);}
\newcommand{\bpthree}{\tikz[baseline=-0.6ex]\filldraw[fill=brown,draw=black] (0,1.1ex)--(1.1ex,0)--(0,-1.1ex)--(-1.1ex,0)--cycle;}
\newcommand{\bpfour}{\tikz[baseline=-0.6ex]\filldraw[fill=red,draw=black] (0,1.2ex)--(1.2ex,-1.0ex)--(-1.2ex,-1.0ex)--cycle;}
\newcommand{\bpfive}{\tikz[baseline=-0.6ex]\filldraw[fill=magenta,draw=black]
(0,1.2ex)--(0.35ex,0.35ex)--(1.2ex,0.35ex)--(0.5ex,-0.15ex)--(0.8ex,-1.1ex)--
(0,-0.45ex)--(-0.8ex,-1.1ex)--(-0.5ex,-0.15ex)--(-1.2ex,0.35ex)--(-0.35ex,0.35ex)--cycle;}
\newcommand{\bpsix}{\tikz[baseline=-0.6ex]\filldraw[fill=cyan,draw=black] (0,0) circle (0.9ex);}
\newcommand{\bpseven}{\tikz[baseline=-0.6ex]\filldraw[fill=orange,draw=black] (-0.9ex,-0.9ex) rectangle (0.9ex,0.9ex);}
\newcommand{\bpeight}{\tikz[baseline=-0.6ex]\filldraw[fill=blue,draw=black] (0,0) circle (0.9ex);}

\title{\boldmath\Large Electro-Weak Phase Transitions and Collider Signals \\
in the Aligned 2-Higgs Doublet Model}
\author[a]{Angela Conaci,}
\author[b]{Stefania De Curtis,}
\author[a]{Luigi Delle Rose,}
\author[c]{Atri Dey,}
\author[a]{Anirban Karan,}
\author[c,d]{Stefano Moretti,}
\author[a]{Maimoona Razzaq}
\affiliation[a]{INFN, Gruppo Collegato di Cosenza \& Dipartimento di Fisica, Università della Calabria, Arcavacata di Rende, I-87036, Cosenza, Italy}
\affiliation[b]{INFN, Sezione di Firenze \& Dipartimento di Fisica e Astronomia, Università di Firenze, Via G. Sansone 1, 50019 Sesto Fiorentino, Firenze, Italy}
\affiliation[c]{Department of Physics and Astronomy, Uppsala University,
Box 516, SE-751 20 Uppsala, Sweden}
\affiliation[d]{School of Physics and Astronomy, University of Southampton, Highfield,
Southampton SO17 1BJ, United Kingdom}
\emailAdd{angela.conaci@unical.it}
\emailAdd{decurtis@fi.infn.it}
\emailAdd{luigi.dellerose@unical.it}
\emailAdd{atri.dey@physics.uu.se}
\emailAdd{anirban.karan@unical.it}
\emailAdd{stefano.moretti@cern.ch}
\emailAdd{maimoona.razzaq@unical.it}
\abstract{We show that the Aligned 2-Higgs Doublet Model (A2HDM) is a framework able to simultaneously 
accommodate strong first order electro-weak phase transitions, in turn generating detectable gravitational waves as well as a variety of Higgs boson signals (involving both the Standard Model  state and its companions, both neutral and charged)  
accessible at the Large Hadron Collider (LHC). We map the
corresponding expanse of parameter space where such a phenomenology is realised in terms of the relative values of the masses of the discovered Higgs boson and the extended Higgs sector states of this model: two neutral ones (a CP-even and a CP-odd) plus a pair of charged ones. We find that both the Laser Interferometer Space Antenna experiment and  High-Luminosity LHC can test such a scenario within their lifetime. This study thus sets the stage for a two-prong complementary approach able to scrutinise the extended Higgs sector of the A2HDM in both its high and low temperature manifestations.    }

\begin{document}

\maketitle

\section{Introduction}

One mystery at the intersection of particle physics and cosmology is the nature of the Electro-Weak Phase Transition (EWPT), the process in the early Universe that activates the Higgs mechanism, in turn giving mass to the fundamental matter fermions and gauge bosons. If such a transition were strongly first order, it could induce perturbations in the primordial plasma capable of generating a stochastic background of Gravitational Waves (GWs) \cite{Kosowsky:1992rz, Kosowsky:1991ua, Kosowsky:1992vn, Nicolis:2003tg, Grojean:2006bp}. In recent years, following the discovery of \bl{a Standard Model (SM)-like} Higgs boson at the Large Hadron Collider (LHC) in 2012 \bl{\cite{ATLAS:2012yve,CMS:2012qbp}}, the birth of Gravitational Wave (GW) cosmology has opened up new ways of probing physics phenomena that involve mass generation, i.e., the transition from the massless unbroken phase to the massive broken one.

Such a strong First-Order EWPT (FOEWPT) cannot be realised within the Standard Model through a Higgs boson with a mass of roughly 125~\cite{Kajantie:1996mn, Kajantie:1996qd, Kajantie:1995kf}. While Beyond the SM (BSM) physics as a means to generate a FOEWPT has been considered long before the Higgs boson mass was determined, it is now especially well motivated by the possibility of probing the corresponding dynamics through GW observations. Furthermore, if such a FOEWPT were to occur over a region of parameter space of some BSM scenario embedding a modified Higgs potential, that would also affect scattering processes at the LHC, thus offering the possibility of establishing a complementarity between collider searches for BSM physics and the quest for additional GW signals of a nature different from those detected by the Laser Interferometer Gravitational-wave Observatory (LIGO) in 2015, which were produced by the merger of two massive black holes~\cite{LIGOScientific:2016aoc}.

One of the simplest extensions of the SM achieving this is the 2-Higgs Doublet Model (2HDM), which, as the name suggests, adds an additional Higgs doublet to the SM while leaving its matter and gauge content unchanged. Investigations of the 2HDM in relation to EWPTs and potentially correlated  collider signals have been pursued for the case of Yukawa structures of Type-I, -II and inert \cite{Goncalves:2021egx, Lee:2025hgb,Wang:2019pet, Benincasa:2022elt, Anisha:2022hgv, Dorsch:2013wja, Basler:2016obg, Bernon:2017jgv, Wang:2018hnw, Su:2020pjw, Andersen:2017ika, Kainulainen:2019kyp, Blinov:2015sna, Hammerschmitt:1994fn, Land:1992sm, Goncalves:2022wbp, Fabian:2020hny,  Fromme:2006cm,  AbdusSalam:2022idz,Aoki:2023lbz, Bittar:2025lcr,Bhatnagar:2025jhh,Anisha:2025zbc,Biermann:2024oyy,Anisha:2023vvu,Basler:2017uxn, Biekotter:2022kgf}.
The purpose of this paper is  to test the possibility that another realisation of the 2HDM, the so-called Aligned 2HDM (A2HDM)~\cite{Pich:2009sp,Coutinho:2024zyp,Karan:2023kyj,Karan:2024kgr,Karan:2023xze,Miralles:2025kes,Coutinho:2024vzm,Braeuninger:2010td,Jung:2010ik, Davila:2025goc,Jung:2013hka,DavilaIllan:2025iwq,Enomoto:2021dkl, Celis:2013rcs, Jung:2010ab,Jung:2012vu,Li:2014fea}, wherein tree level Flavour Changing Neutral Currents (FCNCs) are removed by aligning the Yukawa couplings of the two Higgs doublets in flavour space, thereby avoiding the request of a $Z_2$ symmetry \cite{Pich:2009sp} (as common in other Yukawa realisations), 
can fulfil the dual purpose of generating detectable GWs from strong FOEWPTs, which can eventually be seen by the Laser Interferometer Space Antenna (LISA) experiment~\cite{LISA:2017pwj, Baker:2019nia}, as well as accelerator signals  in scattering experiments accessible at the LHC and/or the High-Luminosity LHC (HL-LHC) \cite{Gianotti:2002xx}.

A further motivation for studying a strong FOEWPT is provided by the baryon asymmetry of the Universe, the clear imbalance between the amount of matter and antimatter in it. \bl{Explaining such an asymmetry through the EW baryogenesis mechanism~\cite{Kuzmin:1985mm,Shaposhnikov:1986jp,Shaposhnikov:1987tw, Cohen:1993nk, Morrissey:2012db, Trodden:1998ym, vandeVis:2025efm} requires the EWPT to be strongly first order, since the resulting departure from thermal equilibrium is one of the necessary Sakharov conditions~\cite{Sakharov:1967dj,Kolb:1990vq}.} As mentioned, the SM does not allow for a FOEWPT, nor does it contain enough CP violation to explain the baryon asymmetry of the Universe. Proposing New Physics (NP) that extends the Higgs sector of the SM is one way to remedy this. In this respect, the A2HDM provides a well-motivated framework in which both the cosmological implications of a strong FOEWPT and its collider phenomenology can be explored.

The plan of this paper is as follows. 
{Sec.~\ref{sec:A2HDM} introduces the tree-level Higgs potential and the complete set of relevant one-loop corrections at both zero and finite temperature, which are essential for the EWPT analysis. The theoretical and experimental constraints on the A2HDM at zero temperature are briefly reviewed in Sec.~\ref{sec:constraints}. The definition of the parameter space is outlined in Sec.~\ref{sec:parameters}. In Sec.~\ref{sec:EWPT&GW}, we perform a detailed study of the FOEWPT and resulting GW signals across the allowed A2HDM parameter space. Sec.~\ref{sec:collider} is devoted to assessing the prospects of probing the A2HDM at the LHC and 
HL-LHC. Our conclusions are presented in Sec.~\ref{sec:conclusion}.} In addition three final Appendices provide further details.

\section{The A2HDM}
\label{sec:A2HDM}

As mentioned, generic 2HDMs~\cite{Branco:2011iw, Gunion:1989we, Ivanov:2017dad} suffer from tree-level FCNCs, which are experimentally constrained to be tiny. In contrast to conventional 2HDMs, the A2HDM eliminates tree-level FCNCs by aligning the Yukawa couplings of the two Higgs doublets in flavour space, without imposing a $Z_2$ symmetry \cite{Pich:2009sp}. Yukawa misalignment arises only at higher-loop orders through minimal flavour-violation and is therefore naturally small, rendering the model radiatively stable~\cite{Penuelas:2017ikk,Braeuninger:2010td,Jung:2010ik}. The A2HDM provides a unified framework encompassing all four conventional 2HDM types \cite{Pich:2009sp} and can also be viewed as a low-energy effective description of more complex scenarios, such as the Composite 2HDM
(C2HDM) 
\cite{DeCurtis:2018iqd,DeCurtis:2018zvh}. Moreover, it allows for additional sources of Charge/Parity (CP) violation that can affect the electron electric dipole moment significantly \cite{Davila:2025goc,Jung:2013hka,DavilaIllan:2025iwq} while remaining compatible with EW baryogenesis \cite{Enomoto:2021dkl}.

\subsection{Model at Tree-Level}

The model describes two complex Higgs doublets with identical hypercharge 1/2. Following EW Symmetry Breaking (EWSB), both fields acquire CP-conserving Vacuum Expectation Values (VEVs). Nevertheless, without loss of generality, it is possible to perform a field redefinition -- corresponding to a rotation in the $SU(2)_L\otimes U(1)_Y$ gauge space -- such that only one of the doublets develops a non-vanishing VEV. This choice defines the so-called \textit{Higgs basis}, in which the Higgs doublets take the form:
\begin{equation}
\Phi_1=\frac{1}{\sqrt 2}\begin{pmatrix}
\sqrt 2\;G^+\\
v+S_1+i\, G^0
\end{pmatrix}\, ,\qquad\qquad \Phi_2=\frac{1}{\sqrt 2}\begin{pmatrix}
\sqrt 2\;H^+\\
S_2+i\, S_3
\end{pmatrix}\, ,
\end{equation} 
with $v=246$~GeV denoting the CP-conserving VEV of $\Phi_1$. The Goldstone modes $G^\pm$ and $G^0$ are used to give masses to the $W^\pm$ and $Z$ bosons, respectively, leaving a physical (pseudo)scalar spectrum comprising of a charged Higgs pair $H^\pm$, two CP-even neutral Higgs states $S_1$ and $S_2$ 
plus a CP-odd neutral particle $S_3$.

The most general 2HDM tree-level scalar potential consistent with the SM gauge symmetry is given by
\begin{align}
\label{eq:pot}
V&=\mu_1^2\,\Phi_1^\dagger \Phi_1+\mu_2^2\,\Phi_2^\dagger \Phi_2+ \Big[\mu_3^2\,\Phi_1^\dagger \Phi_2+\mathrm{h.c.}\Big]+\frac{\lambda_1}{2}\,(\Phi_1^\dagger \Phi_1)^2+\frac{\lambda_2}{2}\,(\Phi_2^\dagger \Phi_2)^2+\lambda_3\,(\Phi_1^\dagger \Phi_1)(\Phi_2^\dagger \Phi_2)\nonumber\\
&+\lambda_4\,(\Phi_1^\dagger \Phi_2)(\Phi_2^\dagger \Phi_1)+\Big[\Big(\frac{\lambda_5}{2}\,\Phi_1^\dagger \Phi_2+\lambda_6 \,\Phi_1^\dagger \Phi_1 +\lambda_7 \,\Phi_2^\dagger \Phi_2\Big)(\Phi_1^\dagger \Phi_2)+ \mathrm{h.c.}\Big],
\end{align}
where $\mu_3^2$, $\lambda_5$, $\lambda_6$ and $\lambda_7$ are, in general, complex parameters. Imposing CP conservation restricts these to be real. The minimisation conditions of the Higgs potential fix $\mu_1^2$ and $\mu_3^2$ in terms of $\lambda_1$ and $\lambda_6$ as follows:
\begin{equation}
\label{eq:minimize}
v^2=-\frac{2\mu_1^2}{\lambda_1}=-\frac{2\mu_3^2}{\lambda_6}\, .
\end{equation}
In the neutral (pseudo)scalar sector, the fields  mix to produce the mass states $h$ and $H$:
\begin{equation}
\label{eq:mix}
\begin{pmatrix}
h\\H
\end{pmatrix}=\begin{pmatrix}
\cos\tilde{\alpha}& \sin\tilde{\alpha}\\ -\sin\tilde{\alpha}&\cos\tilde{\alpha}
\end{pmatrix}\begin{pmatrix}
S_1\\S_2
\end{pmatrix}\, ,
\end{equation}
while $S_3 \equiv A$ remains unmixed. The scalar mass terms derived from the potential can thus be written as
\begin{eqnarray}
\label{eq:mass_pot}
    &V_M= M_{H^\pm}^2\, H^+H^- \, +\, \frac{1}{2}\,M_A^2 A^2\,+\, \frac{1}{2}\begin{pmatrix} S_1 & S_2 \end{pmatrix} \mathcal{M}^2 \begin{pmatrix} S_1 \\ S_2 \end{pmatrix},
    \end{eqnarray}
    where
    \begin{eqnarray}
    \label{eq:massmatrix}
    \hspace{-6mm} M_{H^\pm}^2 = \Big(\mu_2^2+\frac{1}{2}\lambda_3 v^2\Big), \;\;\;\; M_A^2= M_{H^\pm}^2+\frac{1}{2}v^2\left(\lambda_4 - \lambda_5 \right), \;\;\;\; \mathcal{M}^2=\begin{bmatrix} v^2\, \lambda_1  && v^2\, \lambda_6 \\[1mm]
    v^2\, \lambda_6 && M_{H^\pm}^2 +\frac{1}{2}v^2\left(\lambda_4+\lambda_5\right)\end{bmatrix}.
\end{eqnarray}
Diagonalising $\mathcal{M}^2$ gives the squared masses $M_h^2$ and $M_H^2$ of the CP-even mass-eigenstates $h$ and $H$, respectively. Here, the state $h$ is identified with the SM-like Higgs boson with a mass of approximately 125 GeV observed at the LHC.

After eliminating redundant parameters, the Higgs sector of the CP-conserving A2HDM is specified by seven independent quantities: three quartic couplings ($\lambda_2$, $\lambda_3$, $\lambda_7$), the masses of the additional Higgs states ($M_H$, $M_A$, $M_{H^\pm}$) and the mixing angle $\tilde{\alpha}$. The parameters $\mu_1^2$ and $\mu_3^2$ are fixed by Eq.~\eqref{eq:minimize} while the remaining ones can be expressed as: 
\begin{align}
\nonumber  \mu_2^2=M_{H^\pm}^2-&\frac{\lambda_3}{2}v^2\, ,\;\; \lambda_1=\frac{M_h^2+M_H^2\tan^2{\tilde{\alpha}}}{v^2(1+\tan^2{\tilde{\alpha}})}\, ,\;\;  \lambda_4=\frac{1}{v^2}\left( M_h^2+M_A^2-2M_{H^\pm}^2+\frac{M_H^2-M_h^2}{1+\tan^2{\tilde{\alpha}}}\right),\\
     &\lambda_5=\frac{1}{v^2}\left( \frac{M_H^2+M_h^2\tan^2{\tilde{\alpha}}}{1+\tan^2{\tilde{\alpha}}}-M_A^2\right)\, ,\;\;\; \lambda_6=\frac{(M_h^2-M_H^2)\tan{\tilde{\alpha}}}{v^2(1+\tan^2{\tilde{\alpha}})}\,.
\end{align}

In the Higgs basis for the (pseudo)scalar fields and in the mass basis for the  fermion fields, the most general form of Yukawa interactions allowed by the SM gauge symmetry can be expressed as:
\begin{align}
-\mathcal L_Y\, =\, &\bigg(\! 1+\frac{S_1}{v}\!\bigg)\Big\{\bar u_L\,M_u\,u_R+\bar d_L\,M_d\,d_R+\bar l_L\, M_l\,l_R\Big\}\nonumber\\
&+\frac{1}{v}\, (S_2+i S_3) \Big\{\bar u_L\,Y_u\,u_R+\bar d_L\,Y_d\,d_R+\bar l_L\, Y_l\,l_R\Big\}\label{eq:lag_gauge}\\
&+\frac{\sqrt 2}{v}\, H^+ \Big\{\bar u_L\,V\,Y_d\,d_R-\bar u_R \,Y_u^\dagger\,V\,d_L+\bar\nu_L\, Y_l\,l_R\Big\} + \mathrm{h.c.}\, \nonumber
\label{eq:lag_gauge}
\end{align}
Here, $f (\equiv u, d, l)$ denotes the fermions with subscripts $L$ and $R$ indicating Left- and Right-handed chirality, respectively (generation indices have been omitted for brevity). The diagonal mass matrices of the fermions arising from the VEV of $\Phi_1$ are indicated by $M_f$. In contrast, the couplings of the $\Phi_2$ scalar to the fermions are represented by the $3\times3$ matrices $Y_f$ that can induce FCNCs at tree-level. The Cabibbo-Kobayashi-Maskawa (CKM) matrix is denoted by $V$.

The key feature of the A2HDM is the assumption that the matrices $Y_f$ are  proportional to the fermionic mass matrices $M_f$, i.e.,
\begin{equation}
Y_f=\varsigma_f M_f,
\end{equation}
with real-valued alignment parameters $\varsigma_f$ in the CP-conserving scenario. This proportionality ensures that both the matrices are simultaneously diagonalised while suppressing the emergence of FCNCs at tree-level. Consequently, in the mass basis for both fermions and Higgses, the Yukawa interaction Lagrangian reads:
\begin{align}
-\mathcal L_Y = \; & \sum_{f}\bar f M_f \mathcal P_L f+
\sum_{i,f}\;\bigg(\frac{y_f^{\varphi_{0i}}}{v}\bigg)\,\varphi_{0i}\,\Big[\bar f M_f \mathcal{P}_R f\Big] \nonumber \\ 
&+ \bigg(\frac{\sqrt 2}{v}\bigg) H^+\,\Big[\bar u\,\big\{\varsigma_d V M_d \mathcal{P}_R-\varsigma_u M_u^\dagger V\mathcal{P}_L\big\}\, d+\varsigma_l\, \bar \nu M_l \mathcal P_R l\Big] + \mathrm{h.c.} \, ,
\label{eq:lag}
\end{align}
where $\varphi_{0i}$ represents the neutral scalar mass eigenstates ($h$, $H$ and $A$) and $\mathcal{P}_{L,R}$ are the usual chiral projectors. The couplings $y_f^{\varphi_{0i}}$ are determined as follows:
\begin{align}
y_{f}^h=\cos\tilde\alpha+\varsigma_{f}\,\sin \tilde\alpha\, ,  \qquad\quad y_{f}^H=-\sin\tilde\alpha+\varsigma_{f}\,\cos \tilde\alpha\, , \qquad\quad 
y_{u}^A= -i\varsigma_{u}\, , \quad\quad\;  
y_{d,l}^A= i\varsigma_{d,l}\, .
\label{eq:Higgs_yuk}
\end{align}

\bl{(Note that we have chosen the condition of CP-conservation for the sake of simplicity. A more comprehensive analysis incorporating CP-violating effects is left for future investigation.)}

\subsection{Zero Temperature Contributions}
Beyond tree-level, quantum fluctuations of scalar, fermionic and gauge fields modify the Higgs potential through radiative corrections. These effects are incorporated via the one-loop effective potential, which provides a field-dependent description of vacuum stability and (pseudo)scalar interactions.
If $\phi$ denotes the classical background scalar fields, the zero-temperature effective potential is written as
\begin{equation}
\label{eq:veff}
V_{\text{eff}}(\phi)= V_{\text{tree}}(\phi) + V_{\text{CW}}(\phi) + V_{\text{CT}}(\phi).
\end{equation} 
Here, the term $V_{\text{tree}}$ corresponds to the scalar tree-level potential, $V_{\text{CW}}$ defines the Coleman-Weinberg (CW) contribution containing the one-loop radiative corrections and   $V_{\text{CT}}$ denotes the counter-term potential enforcing renormalisation conditions that \bl{ensure that the tree-level minimum is still at least a local minimum of the one-loop potential and preserves the tree-level mass matrix at loop-level}.

In terms of only the CP-even neutral Higgs fields $(\phi_1,\phi_2)$, the two  doublets can be expressed as:
\begin{equation}
    \Phi_i=\frac{1}{\sqrt 2}\begin{pmatrix}
        0\\
        \phi_i
    \end{pmatrix}.
\end{equation}
Thus the tree-level potential in Eq. \eqref{eq:pot} becomes:
\begin{align}
        V_{\rm tree} (\phi_1,\phi_2)= \frac{\mu_1^2}{2}\phi_1^2&+\frac{\mu_2^2}{2}\phi_2^2+ \mu_3^2\phi_1 \phi_2+\frac{\lambda_1}{8}\phi_1^4+\frac{\lambda_2}{8}\,\phi_2^4+\frac{\lambda_{345}}{4}\phi_1^2 \phi_2^2+\frac{\lambda_6}{2}\phi_1^3 \phi_2 +\frac{\lambda_7}{2}\phi_1\phi_2^3
\end{align}
where $\lambda_{345}\equiv (\lambda_3+\lambda_4+\lambda_5)$.

\vspace{3mm}
\noindent
$\bullet$ \textbf{Coleman-Weinberg Potential}\\
In the Modified Minimal Subtraction ($\overline{\text{MS}}$) scheme and Landau gauge, the Coleman-Weinberg contribution~\cite{Coleman:1973jx} reads as 
\begin{equation}
V_{\text{CW}}(\phi)= \sum_k \frac{n_k}{64 \pi^2} \ m^4_k (\phi) \Bigg[ \log\left(\frac{m^2_k (\phi)}{\mu^2}\right) - c_k \Bigg],
\end{equation}
where the index $k$ runs over all particle species contributing to the loop expansion. Here, the quantities in the expression are defined as follows: $n_k$ is the number of degrees of freedom of particle $k$ (negative sign for fermions), $m^2_k (\phi)$ are the field-dependent squared masses obtained by expanding the Lagrangian around the classical scalar background, $\mu$ is the renormalisation scale and $c_k$ denote the scheme-dependent constants equal to $3/2$ for {(pseudo)}scalars and fermions and $5/6$ for gauge bosons.

In the CP-conserving A2HDM, the index $k$ includes the scalar bosons $h, H, A, H^{\pm}$, the Goldstone modes $G^0, G^{\pm}$, the EW gauge bosons $W^{\pm}, Z$ and the SM fermions. The dominant contribution typically arises from the top quark due to its large Yukawa coupling.

The (pseudo)scalar field-dependent masses are obtained from the Hessian of the tree-level potential. Thus, the field-dependent squared-mass-matrices for CP-even ($\mathcal {M}_S^2$), CP-odd ($\mathcal {M}_A^2$) and charged scalars $\mathcal {M}_C^2$ become:
\begin{align}
    \mathcal {M}_S^2(\phi_1,\phi_2) &=\begin{pmatrix}
        \mu_1^2+\frac{3}{2}\lambda_1 \phi_1^2+\frac{1}{2}\lambda_{345} \phi_2^2+3\lambda_6 \phi_1\phi_2 &&& \mu_3^2+\lambda_{345} \phi_1\phi_2+\frac{3}{2}\lambda_6 \phi_1^2+\frac{3}{2}\lambda_7 \phi_2^2\\
        \mu_3^2+\lambda_{345} \phi_1\phi_2+\frac{3}{2}\lambda_6 \phi_1^2+\frac{3}{2}\lambda_7 \phi_2^2 &&& 
        \mu_2^2+\frac{3}{2}\lambda_2 \phi_2^2+\frac{1}{2}\lambda_{345} \phi_1^2+3\lambda_7 \phi_1\phi_2
    \end{pmatrix},\nonumber\\[3pt]
    \mathcal {M}_A^2(\phi_1,\phi_2)&=\begin{pmatrix}
        \mu_1^2+\frac{1}{2}\lambda_1 \phi_1^2+\frac{1}{2}\widetilde\lambda_{345} \phi_2^2+\lambda_6 \phi_1\phi_2 &&& \mu_3^2+\lambda_{5} \phi_1\phi_2+\frac{1}{2}\lambda_6 \phi_1^2+\frac{1}{2}\lambda_7 \phi_2^2\\
        \mu_3^2+\lambda_{5} \phi_1\phi_2+\frac{1}{2}\lambda_6 \phi_1^2+\frac{1}{2}\lambda_7 \phi_2^2 &&& 
        \mu_2^2+\frac{1}{2}\lambda_2 \phi_2^2+\frac{1}{2}\widetilde\lambda_{345} \phi_1^2+\lambda_7 \phi_1\phi_2
    \end{pmatrix},\nonumber\\[3pt]
    \mathcal {M}_C^2(\phi_1,\phi_2)&=\begin{pmatrix}
        \mu_1^2+\frac{1}{2}\lambda_1 \phi_1^2+\frac{1}{2}\lambda_{3} \phi_2^2+\lambda_6 \phi_1\phi_2 &&& \mu_3^2+\frac{1}{2}\lambda_{45} \phi_1\phi_2+\frac{1}{2}\lambda_6 \phi_1^2+\frac{1}{2}\lambda_7 \phi_2^2\\
        \mu_3^2+\frac{1}{2}\lambda_{45} \phi_1\phi_2+\frac{1}{2}\lambda_6 \phi_1^2+\frac{1}{2}\lambda_7 \phi_2^2 &&& 
        \mu_2^2+\frac{1}{2}\lambda_2 \phi_2^2+\frac{1}{2}\lambda_{3} \phi_1^2+\lambda_7 \phi_1\phi_2
    \end{pmatrix},
    \label{eq:field_mass_sc}
\end{align}
where $\widetilde\lambda_{345}\equiv(\lambda_3+\lambda_4-\lambda_5)$ and $\lambda_{45}\equiv(\lambda_4+\lambda_5)$. Upon evaluating the mass matrices at $(\phi_1,\phi_2) = (v,0)$ together with the minimisation condition in Eq.~\eqref{eq:minimize}, the physical mass squared for all scalar degrees of freedom at the EW vacuum are obtained, as shown in Eq. \eqref{eq:massmatrix}.

Furthermore, field-dependent gauge boson masses follow from the kinetic terms of the scalar doublets as:
\begin{equation}
    m_W^{}(\phi_1,\phi_2)=\frac{1}{2}g \sqrt{\phi_1^2+\phi_2^2}\,, \quad m_Z^{}(\phi_1,\phi_2)=\frac{1}{2}\sqrt{g^2+g^{\prime\,2}}\, \sqrt{\phi_1^2+\phi_2^2} \quad \text{and} \quad m_\gamma(\phi_1,\phi_2)=0,
\end{equation}
where $g$ and $g'$ are couplings for $SU(2)_L$ and $U(1)_Y$ gauge interactions respectively. Similarly, field-dependent fermion masses arising from the Yukawa Lagrangian (see Eq. \eqref{eq:lag_gauge}) can be described as:
\begin{equation}
    m_{f_j}(\phi_1,\phi_2)=\frac{(M_f)_{jj}}{v}(\phi_1+\varsigma_f\,\phi_2),
\end{equation}
where $M_f$ denotes the diagonal mass matrix of the fermions and $j$ indicates the generation.

\vspace{3mm}
\noindent
$\bullet$ \textbf{Counter-term Potential and Renormalisation Conditions}\\
The inclusion of radiative corrections shifts both the VEV and the (pseudo)scalar masses from their corresponding tree-level values. To ensure that the VEV and masses remain unchanged at one-loop order compared to their tree-level values, a counter-term potential $V_{\rm CT}$ is introduced  by imposing renormalisation conditions on the first and second derivatives of the one-loop effective potential $V_{\rm eff}$:
\begin{equation}
\left.\partial_i V_{\rm eff}\right|_{v} = 0 \qquad \text{and} 
\qquad
\left.\partial_i \partial_j V_{\rm eff}\right|_{v}
=
\left.\partial_i \partial_j V_{\rm tree}\right|_{v},
\end{equation}
with $(i,j)=\{1,2\}$ and $\partial_i \equiv {\partial}/{\partial \phi_i}$.
Since the tree-level potential, $V_{\rm tree}$ already satisfies these conditions, the first and second derivatives of the counter-terms must cancel  similar contributions from the CW potential, i.e.:
\begin{equation}
\label{eq:renorm}
\left.\partial_i V_{\rm CT}\right|_{v}
=
\,-\left.\partial_i V_{\rm CW}\right|_{v}
\qquad \text{\rm and } \qquad
\left.\partial_i \partial_j V_{\rm CT}\right|_{v}
=
\,-\left.\partial_i \partial_j V_{\rm CW}\right|_{v},
\qquad (i,j=1,2).
\end{equation}
The counter-term potential then reads,
\begin{equation}
\label{eq:CT_pot}
V_{\mathrm{CT}}
=
\delta m_1^2 \phi_1^2
+
\delta m_3^2 \phi_1 \phi_2
+
\delta\lambda_1 \phi_1^4
+
\delta\lambda_{345} \phi_1^2 \phi_2^2
+
\delta\lambda_6 \phi_1^3 \phi_2.
\end{equation}
In principle, one may introduce a more general structure for it~\cite{Cline:2011mm}. However, since CP violation and charge-breaking minima are not considered in the present analysis, it is sufficient to restrict the counter-term potential to the form given above. We emphasise that our $V_{\mathrm{CT}}$ differs from those commonly employed in the literature for 2HDM~\cite{Aoki:2021oez, Bernon:2017jgv}. This difference originates from our choice of the \textit{Higgs basis}, in which the renormalisation conditions are imposed at $(v,0)$.

The counter-term potential in Eq.~\eqref{eq:CT_pot} contains five unknown parameters\footnote{Although the counter-term potential  could possibly include another term of the form $\delta m_2^2\, \phi_2^2$, we set $\delta m_2^2 = 0$. This choice does not affect the vacuum configuration at zero temperature, as $\phi_2$ does not acquire a VEV in the Higgs basis.}, which can be uniquely determined by imposing the five renormalisation conditions given in Eq.~\eqref{eq:renorm}. This leads to:
\begin{align}
\hspace*{-5mm}\delta m_1^2 =
\frac{1}{4}\bigg(\partial_1^2\, & V_{\rm CW}
-\frac{3}{v}\partial_1 V_{\rm CW}\bigg),\quad
\delta m_3^2 =
\frac{1}{2}\bigg(\partial_1\partial_2 V_{\rm CW}
-\frac{3}{v}\partial_2 V_{\rm CW}\bigg), \quad
\delta\lambda_{345}
=
-\,\frac{1}{2v^2}\,\partial_2^2\, V_{\rm CW},\nonumber\\[2pt]
&\hspace*{-5mm}\delta\lambda_1 =
\frac{1}{8v^2}
\bigg(
-\partial_1^2\, V_{\rm CW}
+\frac{1}{v}\partial_1 V_{\rm CW}
\bigg),\quad 
\delta\lambda_6 =
\frac{1}{2v^2}
\bigg(
-\partial_1\partial_2 V_{\rm CW}
+\frac{1}{v}\partial_2 V_{\rm CW}
\bigg).
\label{eq:CTs}
\end{align}
Thus, the counter-terms are estimated by evaluating the derivatives of the CW potential at zero temperature with the inclusion of the Infrared Regularised (IR) Goldstone contributions. To avoid the IR divergences in the second derivatives of $V_{\text{CW}}$, which depend on the logarithm of the Goldstone boson masses, we rely on the approximation found in \cite{Cline:2011mm}, where an IR cut-off has been introduced and taken to be equal to the mass of the SM-like Higgs boson. 

\subsection{Finite Temperature Contributions}
To study the thermal evolution of the scalar sector and investigate the  EWPTs, the effective potential must be extended to finite temperature. At one-loop level, the finite temperature effective potential reads
\begin{equation}
V_{\text{eff}}(\phi, T)= V_{\text{eff}}(\phi) +
V_{\text{T}}(\phi, T) +
V_{\text{daisy}}(\phi,T),
\end{equation}
where $V_{\text{eff}}(\phi)$ is the zero-temperature effective potential (see Eq. \eqref{eq:veff}), $V_T$ accounts for the one-loop thermal corrections and $V_{\text{daisy}}$ implements daisy resummation needed to handle IR divergences in bosonic thermal loops.\\
\\
$\bullet$ \textbf{One-loop Thermal Contributions}\\
The finite temperature one-loop contribution is
\begin{equation}
V_{T}(\phi, T) = \frac{T^4}{2 \pi^2} \sum_k n_k J_{\pm}\left( \frac{m^2_k (\phi)}{T^2}\right),
\end{equation}
where the thermal functions $J_{\pm}$ are defined as
\bl{\begin{equation}
    J_{\pm}(y^2)= \int^{\infty}_{0} dx \ x^2 \ln \left[ 1 \pm e^{- \sqrt{x^2 + y^2}}\right],
\end{equation}
with the upper sign for fermions and the lower sign for bosons.} As already mentioned, $n_k$ is positive for bosons and negative for fermions. At high temperatures $(T \gg m_k)$, these functions can be expanded in series of $m_k/T$, i.e.
\begin{align}
\bl{J_+(y^2)} &\bl{\,\approx \frac{7 \pi^4}{360} - \frac{\pi^2}{24}\, y^2 - \frac{1}{32}\, y^4 \ln \frac{y^2}{a_f} + \dots,} \\
J_-(y^2) &\,\approx -\frac{\pi^4}{45} + \frac{\pi^2}{12}\, y^2 - \frac{\pi}{6}\, y^3 - \frac{1}{32}\, y^4 \ln \frac{y^2}{a_b} + \dots
\end{align}
where $a_b$ and $a_f$ are constants arising from the renormalisation of thermal integrals \cite{Anderson:1991zb, Quiros:1999jp}. \\ \\
\bl{Furthermore, in the low temperature limit ($T\ll m_k$),  these integrals $J_\pm$ can be approximated using the asymptotic expansion of Bessel-function in the following form~\cite{Anderson:1991zb}:
\begin{equation}
    J_\pm(y^2)\approx\pm \sqrt{\frac{\pi}{2}}\, y^{3/2}\, e^{-y} \,\Big(1+\frac{15}{8y}+\cdots\Big) \,.
\end{equation}} \\
$\bullet$ \textbf{Daisy Resummation}\\
At high temperatures, bosonic infrared divergences become significant due to the presence of massless modes in the thermal bath. These divergences arise from zero-Matsubara-frequency modes  
in the thermal propagators, rendering naive perturbation theory unreliable. To cure this, daisy resummation \cite{Arnold:1992rz, Gross:1980br, Parwani:1991gq} sums the leading thermal contributions from diagrams with an arbitrary number of ``ring'' insertions (bosonic self-energy loops). This effectively replaces the tree-level masses in the thermal loops by the thermal masses
\begin{equation}
m^2_k(\phi) \to \tilde{m}^2_k(\phi,T)=m^2_k(\phi) + \Pi_k(T), 
\end{equation}
where $\Pi_k(T)$ is the finite temperature self-energy of a particle $k$, typically proportional to 
$T^2$.

Two schemes are used: the one by Arnold–Espinosa   \cite{Arnold:1992rz}, where thermal masses are included only in the cubic term of the high-temperature expansion, and the Parwani one, where thermal masses are included in the full one-loop functions replacing $m_k \to \tilde{m}_k (T)$ \cite{Parwani:1991gq}. In our work, we used the Parwani scheme. \bl{This prescription offers a convenient implementation of thermal screening in the numerical evaluation of the finite-temperature effective potential, since the thermally corrected masses are used directly in the one-loop thermal functions. The Arnold-Espinosa prescription provides an alternative resummation scheme, differing from the Parwani one by terms that are formally beyond the perturbative order considered here.}

The thermal coefficients introduce thermal corrections to the squared-mass-matrices for the states  $\mathcal {M}_S^2$, $\mathcal {M}_A^2$ and $\mathcal {M}_C^2$ (see Eq. \eqref{eq:field_mass_sc}) as:
\begin{equation}
\mathcal {M}_X^2 (\phi_1,\phi_2,T) \to \mathcal {M}_X^2(\phi_1,\phi_2) + 
\begin{pmatrix} c_1 & c_3 \\ c_3 & c_2 \end{pmatrix} T^2, \qquad  X \in  \{S,A,C\}, 
\end{equation}
and the thermally corrected field-dependent scalar masses are obtained by diagonalising these matrices.

The thermal coefficients for the scalar fields are given by:
\begin{align}
c_1 &= \frac{g^2}{8} + \frac{g^2 + g^{\prime 2}}{16} + \frac{\lambda_1}{4} + \frac{\lambda_3}{6} + \frac{\lambda_4}{12} 
      + \frac{\tilde y_t^2}{4} + \frac{\tilde y_b^2}{4} + \frac{\tilde y_\tau^2}{12}, \\
c_2 &= \frac{g^2}{8} + \frac{g^2 + g^{\prime 2}}{16} + \frac{\lambda_2}{4} + \frac{\lambda_3}{6} + \frac{\lambda_4}{12} 
      + \frac{\varsigma_u^2\, \tilde y_t^2}{4} + \frac{\varsigma_d^2\, \tilde y_b^2}{4} + \frac{\varsigma_l^2\, \tilde y_\tau^2}{12}, \\
c_3 &= \frac{\lambda_6}{4} + \frac{\lambda_7}{4} 
      + \frac{\varsigma_u\, \tilde y_t^2}{4} + \frac{\varsigma_d\, \tilde y_b^2}{4} + \frac{\varsigma_l\, \tilde y_\tau^2}{12},
\end{align}
where, $\tilde{y}_{f}=(\sqrt 2 m_f/v)$ with $m_f$ being the physical fermion mass at zero temperature (we are neglecting the contributions from light fermions). It is worth emphasising that these thermal coefficients differ from those commonly reported in the literature for the 2HDM Type-I and -II \cite{Benincasa:2022elt,Aoki:2021oez} where only one of the scalar doublets couples to the fermion fields (in the gauge basis) in order to preserve the $Z_2$ symmetry. Consequently, only one of the coefficients, either $c_1$ or $c_2$, receives contributions from Yukawa interactions. In contrast, in the present framework both $c_1$ and $c_2$ depend on the Yukawa couplings. Since the one-loop thermal self-energy of $\Phi_2$ involves two $\bar f f \phi_2$ vertices, the corresponding contributions to $c_2$ are proportional to $\varsigma_f^2$. 

Furthermore, unlike in the other 2HDMs, thermal corrections to $\mu_3^2$ arise in this setup. At one-loop order, these corrections do not receive contributions from gauge bosons, as the kinetic terms do not induce mixing between $\phi_1$ and $\phi_2$. In the Yukawa sector, the corresponding corrections are proportional to a single power of $\varsigma_f$.

For the gauge bosons, the resummation affects only the longitudinal components. 
The thermally corrected squared-mass of the longitudinal $W^\pm$ boson is
\begin{equation}
m^2_{W_L^\pm} (\phi_1,\phi_2,T) = \frac{g^2}{4} \left( \phi_1^2 + \phi_2^2 \right) + 2 g^2 T^2,
\end{equation}
while the  thermally corrected squared-mass matrix involving the longitudinal components of $W_3$  and $B$ bosons reads:
\begin{equation}
\frac{1}{4} \left( \phi_1^2 + \phi_2^2 \right)
\begin{pmatrix} g^2 & -g g' \\ -g g' & g'^2 \end{pmatrix} +
\begin{pmatrix} 2 g^2 T^2 & 0 \\ 0 & 2 g'^2 T^2 \end{pmatrix}.
\end{equation}
Upon diagonalisation this matrix gives the thermally corrected squared-masses for longitudinal components of the $Z$ boson and of the photon as:
\begin{align}
m^2_{Z_L} (\phi_1,\phi_2,T)&= \frac{1}{8} (g^2 + g'^2) \left( \phi_1^2 + \phi_2^2 + 8 T^2 \right) + \Delta, \\
m^2_{\gamma_L} (\phi_1,\phi_2,T)&= \frac{1}{8} (g^2 + g'^2) \left( \phi_1^2 + \phi_2^2 + 8 T^2 \right) - \Delta,
\end{align}
where
\begin{equation}
\Delta^2 = 
\left[ \frac{1}{8} (g^2 + g'^2) (\phi_1^2 + \phi_2^2 + 8 T^2) \right]^2
- g^2 g'^2 T^2 (\phi_1^2 + \phi_2^2 + 4 T^2).
\end{equation}

\section{Constraints}
\label{sec:constraints}

In this section we summarise the constraints on the zero-temperature parameter space of the A2HDM. Detailed discussions of these constraints can be found in Refs.~\cite{Coutinho:2024zyp,Karan:2023kyj,Karan:2024kgr,Karan:2023xze,Miralles:2025kes,Coutinho:2024vzm}.

\subsection{Vacuum Stability, Perturbativity and Unitarity}
To ensure \bl{the absolute stability of the vacuum} at zero temperature, the Higgs potential must be bounded from below and it is also necessary to verify that the ground state of the potential corresponds to a global minimum. \bl{Although a (sufficiently) metastable local minimum (e.g., as in the SM) may present a valid zero-temperature ground state, we restrict our analysis to the region of parameter space with such a absolute vacuum stability.} Deriving analytical conditions for vacuum stability in the A2HDM is significantly more complicated than in the usual 2HDM scenarios due to the presence of the $\lambda_6$ and $\lambda_7$ terms in the scalar potential. We employ here the so-called \textit{bilinear formalism}, as described in Refs.~\cite{Ivanov:2006yq,Ivanov:2015nea,Bahl:2022lio}, in which the potential is expressed in terms of bilinear combinations of the Higgs doublets to impose conditions for a stable vacuum.

In order to maintain perturbativity within the scalar sector, the quartic couplings are constrained as $|\lambda_i|< 4\pi$. Analogously, in the Yukawa sector we impose the condition that the couplings of fermions to the charged Higgs states remain below unity, i.e., $|\varsigma_f|< v/(\sqrt{2} m_f)$. 

Furthermore, we require \textit{perturbative unitarity} to hold, such that none of the $2 \to 2$ scalar scattering matrix elements diverges with increasing Center-of-Mass (CoM) energy. Denoting 
the tree-level scattering amplitude matrix by $\mathbf{a_0}$, with 
$a_j^0$ the eigenvalue in the $j^{\text{th}}$ partial wave, we note that at high energies the dominant contribution arises from the $S$-wave ($j=0$). It is therefore sufficient to restrict attention to this case. The resulting condition for tree-level unitarity is then expressed as
\begin{equation}
\label{eq:unitarity}
    (a_0^{0})^2\leq \frac{1}{4} \qquad\text{with}\qquad (\mathbf{a_0})_{i,f}=\frac{1}{16\pi s}\int_{-s}^{0} dt \;\mathcal M_{i\to f}(s,t)\, .
\end{equation}
In this framework, one obtains fourteen neutral, eight singly charged and three doubly charged two-body scattering processes involving Higgs states. For practical purposes, the $25$-dimensional scattering matrix $\mathbf{a_0}$ is typically block-diagonalised in the basis of hypercharge and weak isospin of the scattering states \cite{Ginzburg:2005dt,Bahl:2022lio}. (A similar approach, formulated in a slightly different basis, is presented in Ref. \cite{Kanemura:2015ska}.)

To ensure the reliability of the perturbative expansion, we further require the one-loop vertex corrections to the cubic coupling $\varphi^0_i H^+H^-$ to remain below $50\%$  their corresponding tree-level values, as suggested in Ref. \cite{Celis:2013rcs}. This interaction is particularly relevant in scenarios with light charged Higgs states and also plays a significant role in the decays $\varphi^0_i \to \gamma\gamma$.

\subsection{Oblique Parameters}

The vacuum polarisation of the EW gauge bosons receives contributions from the extended Higgs sector of the A2HDM, leading to modifications of the gauge boson propagators. These effects are conveniently parametrised by the oblique parameters $S$, $T$ and $U$, also known as the Peskin–Takeuchi variables  \cite{Peskin:1990zt,Peskin:1991sw}. In the A2HDM, the contribution to $U$ is negligible \cite{Haber:2010bw} and therefore only $S$ and $T$ are employed as observables in our fit. It should be emphasised that we adopt the values of the oblique parameters uncontaminated from the (pseudo)scalar-induced effects, as specified in Ref. \cite{Karan:2023kyj}. These values are extracted from a global EW fit where the observable $R_b\equiv \Gamma(Z\to b\bar b) / \Gamma(Z\to \text{hadrons})$ is excluded, since it also receives (pseudo)scalar-induced corrections within the A2HDM \cite{Haber:1999zh,Degrassi:2010ne}.

\subsection{Flavour Observables}

In our analysis, we take into account all flavour observables relevant to CP-conserving NP scenarios. The set includes loop-induced processes such as $\Delta M_{B_s}$~\cite{Jung:2010ik,Chang:2015rva} and the following Branching Ratios (BRs): ${\rm{BR}}(B \to X_s \gamma)$~\cite{Bobeth:1999ww, Misiak:2006ab, Misiak:2006zs, Jung:2010ik, Jung:2010ab, Jung:2012vu, Hermann:2012fc, Misiak:2015xwa, Misiak:2017woa, Misiak:2020vlo}, ${\rm{BR}}(B_s \to \mu^+ \mu^-)$~\cite{Li:2014fea,Arnan:2017lxi}, $R_b$ and the muon $(g-2)\mu$~\cite{Ilisie:2015tra,Lautrup:1971jf, Leveille:1977rc, Czarnecki:1995wq, Dedes:2001nx, Chang:2000ii, Cheung:2001hz, Cheung:2003pw,  Cherchiglia:2016eui, Athron:2021evk,Aliberti:2025beg}. We also incorporate observables associated with tree-level transitions~\cite{Jung:2010ik, ParticleDataGroup:2024cfk}, including ${\rm{BR}}(B \to \tau \nu)$, ${\rm{BR}}(D{(s)} \to \mu \nu)$, ${\rm{BR}}(D_{(s)} \to \tau \nu)$, as well as the ratios $\Gamma(K \to \mu \nu)/\Gamma(\pi \to \mu \nu)$ and $\Gamma(\tau \to K \nu)/\Gamma(\tau \to \pi \nu)$. All these flavour observables are affected by the presence of the additional Higgs states. It should be further noted that the CKM matrix elements are determined through separate fits to flavour observables that remain unaffected by the presence of additional (pseudo)scalar contributions \cite{Karan:2023kyj}.

\subsection{Higgs Signal Strengths}

The ATLAS and CMS collaborations have investigated the dominant production mechanisms, gluon-gluon Fusion (ggF), Vector Boson Fusion (VBF), associate production with a vector boson ($Vh$)  and a top-quark pair ($t\bar{t}h$), combined with the decay channels $h\to c\bar{c}$, $b\bar{b}$,  $\gamma\gamma$, $\mu^+\mu^-$, $\tau^+\tau^-$, $W^+W^-$ and $Z\gamma$ and $ ZZ $ in considerable detail. Within the framework of the A2HDM, modifications to the couplings of the SM-like Higgs boson with weak gauge bosons and fermions induce deviations in the Higgs signal strengths, defined as the production cross section for a given mode multiplied by the BR into a specific final state, normalised to the corresponding SM prediction. In this work, we employ the 13 TeV data reported by ATLAS~\cite{ATLAS:2022vkf} and CMS~\cite{CMS:2022dwd}, together with their combined results at 8 TeV~\cite{ATLAS:2016neq}. (Notice that the Higgs decay to a charm-quark pair, included in more recent experimental analyses~\cite{CMS:2022psv,ATLAS:2022ers}, is also taken into account in our fit).

\subsection{Direct Searches}

For direct searches of new Higgs states, we make use of roughly 150 experimental results from ATLAS and CMS (at both 8 TeV and 13 TeV) as well as from LEP. Below we provide a brief overview of the relevant searches: for brevity, full decay channels are not listed here but can be found in detail in Ref. \cite{Coutinho:2024zyp}.

For resonant production of a heavy neutral Higgs boson (with mass larger than $m_h$), we include ATLAS and CMS searches for its decays into the following final states: fermion pairs, gauge boson pairs, Higgs pairs, $hZ$ and a $Z$ boson associated with another lighter neutral scalar. For light scalars (with mass smaller than $m_h$), ATLAS and CMS mainly present results on resonant $h$ production followed by decays into pairs of lighter neutral scalars and pseudoscalars, which we also use. In addition, we incorporate ATLAS and CMS searches for the associated production of light scalars with a 
top-quark pair, bottom-quark pair and a weak gauge boson, along with the $pp \to H \to \gamma\gamma$ and $pp \to A \to \tau^+\tau^-$ channels. LEP searches for pair production of light neutral (pseudo)scalars and for associated production of a $Z$ boson with a light neutral pseudoscalar, both via Higgsstrahlung, are also included.

For heavy charged Higgs states, and $M_{H^\pm}>m_t$, we consider ATLAS and CMS results on resonant production followed by decays into $\tau^+\nu$ and $t\bar b$. For light charged Higgs states, we include ATLAS and CMS searches for the top-quark decay channel $t \to H^+ b$ as well as LEP limits on $H^+H^-$ production via  $\gamma,Z$ mediation in $s$-channel.

\subsection{Other Observables}

In addition to the total width of the top quark, we incorporate the invisible widths of the $W^\pm$, $Z$ and Higgs bosons, which are well measured experimentally \cite{ParticleDataGroup:2024cfk} and will be modified in the presence of light Higgs bosons' states. We also account for LHC and LEP searches for supersymmetric sleptons (see Ref. \cite{Coutinho:2024zyp} for details), since their decays into a charged lepton and a neutralino yield collider signatures that closely resemble those of charged Higgs decays into a charged lepton and a neutrino.

\section{A2HDM Parameter Space}
\label{sec:parameters}

For generation of parameter points compatible with all the constraints mentioned above we use the open-source \texttt{HEPfit} package~\cite{DeBlas:2019ehy} that implements a Markov-Chain Monte-Carlo (MCMC)  algorithm via the \texttt{Bayesian Analysis Toolkit}~\cite{Caldwell:2008fw} and perform global fits of the A2HDM. We begin with flat distributions of model parameters as \textit{priors}. Then, based on the experimental measurement of a large number of observables, \texttt{HEPfit} constructs the \textit{likelihood} $(\mathscr L)$ of the model, according to Bayes' theorem, thereby generating the \textit{posterior} distribution. The relative quality of different fits or models is assessed using the Information Criterion (IC), defined as ${\rm IC}= -2\,\overline{\log \mathscr L} + 4\,\sigma^2_{\log \mathscr L}$, where $\overline{\log \mathscr{L}}$ denotes the mean of the log-likelihood function and $\sigma^2_{\log \mathscr{L}}$ its variance. A smaller IC value indicates a better performing fit to a model.

\begin{table}[h!]
    \centering
\begin{tabular}{P{1.1cm}|P{1.1cm}|P{1.1cm}|P{1.1cm}|P{1.1cm}|P{1.1cm}|P{1.1cm}|P{1.1cm}|P{1.1cm}|P{1.1cm}|P{1.1cm}|P{1.1cm}|}
\hline
\multicolumn{12}{|c|}{\bf Priors} \\
\hline
\hline
\multicolumn{6}{|P{6.6 cm}}{$M_{\phi_{\text{light}}^{}} \in [10 \text{ GeV},\; M_h]$ } & \multicolumn{6}{|P{6.6 cm}|}{$M_{\phi_{\text{heavy}}^{}} \in [M_h,\; 1 \text{ TeV}]$ }  \\
\cline{2-12}
\multicolumn{4}{|P{4.4 cm}}{$\lambda_2\in [-1,\;10]$} & \multicolumn{4}{|P{4.4 cm}|}{$\lambda_3\in [-1,\;4\pi]$}  & \multicolumn{4}{P{4.4 cm}|}{$\lambda_7\in [-3.5,\;3.5]$}  \\
\cline{2-12}
\multicolumn{3}{|P{3.3 cm}}{$\tilde \alpha\in [-0.2,\;0.2]$} & \multicolumn{3}{|P{3.3 cm}|}{$\varsigma_u\in [-1.5,\; 1.5]$} & \multicolumn{3}{P{3.3 cm}|}{$\varsigma_d \in [-25,\; 25]$} & \multicolumn{3}{P{3.3 cm}|}{$\varsigma_l \in [-100,\; 100]$} \\  
\hline
    \end{tabular}
    \caption{Chosen priors for the parameters of the A2HDM. Here,  $\phi_{\text{light}}^{}$ and $\phi_{\text{heavy}}^{}$ indicate Higgs states  lighter and heavier than the SM-like Higgs boson, respectively, with $\lbrace\phi_{\text{light}}^{},\; \phi_{\text{heavy}}^{}\rbrace\in \lbrace H,\; A,\; H^\pm\rbrace$.}
    \label{tab:prior}
\end{table}

It has been shown in Refs.~\cite{Coutinho:2024zyp,Karan:2023kyj} that the A2HDM framework accommodates both heavy and light Higgs states. The full parameter space can be systematically explored by dividing it into eight distinct regions, classified according to whether the additional Higgs masses lie above or below the measured Higgs mass value, $M_h=125.2$ GeV. These comprise: (i) a scenario with all extra Higgses heavier than $M_h$; (ii) three cases with a single lighter state; (iii) three cases with two lighter states;  (iv) a scenario where all the extra Higgses are lighter than $M_h$. Flat priors for the MCMC scans in \texttt{HEPfit} are adopted as listed in Tab.~\ref{tab:prior}, with ranges chosen to ensure complete coverage of the physically relevant parameter space for each scenario. 

Using {\tt HEPfit}, we generate a large number of parameter points for all eight scenarios, as dictated by the combined likelihood incorporating all relevant constraints. Each of these points is then passed through {\tt CosmoTransitions}~\cite{Wainwright:2011kj} to evaluate EWPT related quantities, which are subsequently used to predict the corresponding GW signatures.  (Notice that our results have been verified against {\tt BSMPT}~\cite{Basler:2018cwe, Basler:2020nrq, Basler:2024aaf}). We will then select those parameter points that yield a large Signal-to-Noise Ratio  (SNR) for GW detection and subsequently examine their testability at HL-LHC \cite{Gianotti:2002xx}. The production cross sections at the CERN machine for various channels involving the BSM Higgs states are computed using {\tt MadGraph5\_aMC\@NLO}~\cite{Alwall:2014hca,Frederix:2018nkq}. \\
 
 As shown in Ref.~\cite{Karan:2023kyj} and from the comparison of the IC values in Tab.~\ref{tab:IC}, the scenarios in which either all the extra Higgses are heavy or only one is light show a better fit to the data, whereas the remaining scenarios exhibit a somewhat poorer performance. The masses of the BSM Higgs states are primarily constrained by requirements of vacuum stability and perturbativity and from direct searches at colliders. At 95\% Confidence Level (CL), LHC searches for $h \to AA$ and $h \to HH$ limit the new neutral Higgs masses to be above 62.5~GeV (i.e., $M_h/2$) while LEP searches require the charged Higgs state to be heavier than about 90~GeV. The mass splittings among the additional Higgs states are governed by the oblique parameters, which, in scenarios containing light (pseudo)scalars, restrict the heaviest Higgs masses to be below 600~GeV (at 95\% CL). In the scenario where all additional BSM Higgs states are heavy, no observable imposes an upper limit on their masses, however, we restrict the priors up to 1~TeV to ensure efficient sampling of the parameter space in the range [$M_h$, 1~TeV]. The quartic couplings $\lambda_2$, $\lambda_3$ and $\lambda_7$ are constrained by the stability and perturbativity conditions. Among these, $\lambda_2$ remains within the range [0,8.5] in all scenarios, whereas the posterior ranges of $\lambda_3$ and $\lambda_7$ depend on the masses of $H^\pm$ and $H$, respectively\footnote{The posterior range of $\lambda_7$ is largely determined by the trilinear $hH^+H^-$ coupling, which contributes to the $h \to \gamma\gamma$ decay.}, therefore varying between different scenarios. The measured Higgs signal strengths constrain the mixing angle $\tilde{\alpha}$ to values close to zero and confine all the scenarios within the domains bounded by four hyperbolas in the $(\tilde\alpha$~-~$\varsigma_f)$  plane\footnote{The fermionic decay modes of the Higgs signal strengths constrain  $|y_f^h|$ (see Eq.~(\ref{eq:Higgs_yuk})). The experimental uncertainties define four hyperbolic boundaries in the $(\tilde\alpha$~-~$\varsigma_f)$ plane. The opposite-sign solutions in $\varsigma_u$ and $\varsigma_d$ are disfavoured by direct search constraints.}. Flavour observables further restrict the alignment parameters $\varsigma_u$ and $\varsigma_d$, depending on the $H^\pm$ mass, since the new physics contributions scale as quadratics in $(\varsigma_u/M_{H^\pm})$ and $(\varsigma_d/M_{H^\pm})$. In contrast, direct searches, particularly those involving charged Higgs states, constrain $\varsigma_l$ through bounds on the products $\varsigma_l \varsigma_u$ and $\varsigma_l \varsigma_d$. Regarding slepton searches, only the LHC searches for $\tilde{\tau}_L$ have a mild impact on the A2HDM parameter space.

\begin{table}[h!]
    \hspace{-5mm}
    \begin{tabular}{|c|c|c|c|c|c|c|c|}
    \hline
      All Heavy & Light A& Light H& Light H$^\pm$& Light A\&H$^\pm$ &Light H\&H$^\pm$& Light A\&H& All Light \\
      \hline 
       83.70&84.34&84.93&87.51&87.61&88.46&90.65&91.27 \\
       \hline
    \end{tabular}
    \caption{The IC values corresponding to the fits in different regions of parameter space. A smaller IC value indicates a better performing fit.}
    \label{tab:IC}
\end{table}

\section{EWPT Dynamics and GW Signals}
\label{sec:EWPT&GW}

The extended Higgs sector of the A2HDM allows the EWSB to proceed through strong first-order transitions. Such transitions are expected to leave an observable imprint in the form of a stochastic background of GWs. In this section, we derive the corresponding spectrum for all of the eight scenarios considered and assess their detectability by the LISA experiment. 

\subsection{EWPT Parameters and Bubble Dynamics}

As the Universe expanded and cooled, the potential energy landscape of (pseudo)scalar fields underwent significant changes. At extremely high temperatures, thermal contributions dominated, keeping the fields stabilised near zero. As the temperature drops, the potential develops additional minima at finite field values, creating metastable configurations. In cases where the transition is first order, the system can move from a metastable (false vacuum) to a stable (true vacuum) state through the nucleation of bubbles, which grow and merge. This process can produce stochastic GWs.

The bubble nucleation rate per unit volume can be expressed as \cite{Linde:1980tt, Linde:1981zj}
\begin{equation}
\Gamma(T) \sim A(T) \, e^{-S_3(T)/T},
\end{equation}
where $A(T)$ is a prefactor with dimensions of $(\text{energy})^4$ and $S_3(T)$ is the three-dimensional Euclidean action evaluated for a critical bubble configuration with spherical $O(3)$ symmetry:
\begin{equation}
S_3(T) = 4 \pi \int_0^\infty dr \, r^2 \left[ \frac{1}{2} \left( \frac{d\phi}{dr} \right)^2 + V_{\rm eff}(\phi,T) - V_{\rm eff}(\phi_{\rm false},T) \right].
\end{equation}
Bubble nucleation occurs when the probability of forming critical bubbles becomes sizeable, defining the nucleation temperature $T_n$, at which 
\begin{equation}
\Gamma(T_n) H(T_n)^{-4} \sim 1.
\end{equation}
Here, $H(T_n)$ is the Hubble parameter at $T_n$. This temperature marks the onset of the EWPT.

During a strong FOEWPT, the probability for the Universe to nucleate bubbles of the true vacuum depends on the shape of the potential barrier separating the metastable and stable phases. This same potential difference that governs the nucleation rate also determines how much energy is released. The strength of the EWPT is usually quantified by the ratio of released vacuum energy to the energy density of the surrounding radiation \cite{Espinosa:2010hh}
\begin{equation}
    \alpha =\frac{\Delta \rho_{\text{vac}}(T_n)}{\rho_{\text{rad}} (T_n)},   \ \ \ \ \ \ 
\text{with} \ \ \ \ \ \rho_{\text{rad}}(T_n)= \frac{\pi^2}{30} g_* (T_n) \ T_n^4,
\end{equation}
where $\Delta \rho_{\text{vac}}= \Delta V_{\text{eff}} \ - \frac{T}{4} \Delta (\frac{\partial V_{\text{eff}}}{\partial T})$ and $g_*$ is the effective number of relativistic degrees of freedom at the temperature of the transition.  The parameter $\beta/H$ \cite{Grojean:2006bp} describes instead the inverse duration of the transition relative to the Hubble rate, which is given by 
\begin{equation}
    \frac{\beta}{H}
= T_n \left. \frac{d}{dT} \left( \frac{S_3(T)}{T} \right) \right\rvert_{T = T_n}.
\end{equation}
The features of the GW spectrum are mainly determined by the interplay of $\alpha$ and $\beta/H$ \footnote{For sufficiently fast EWPTs ($\beta/H \gg 1$), 
 as we shall see below happening for our BSM scenario, the nucleation and percolation temperatures nearly coincide, i.e., $T_n \approx T_p$. In our parameter space, then, this justifies using $T_n$ in the computation of the GW signal.}. Stronger transitions (larger $\alpha$) produce more intense GW signals, while $\beta/H$ controls the spectrum frequency and width.

\subsection{GW Spectrum}
 When the bubbles nucleated during the EWPT expand and collide, a fraction of the latent heat is converted into anisotropic stresses in both the scalar field and the plasma, generating a stochastic GW background. The present-day energy density receives contributions from bubble collisions $\Omega_{\text{col}}$, sound waves in the plasma $\Omega_{\text{sw}}$ and turbulence $\Omega_{\text{turb}}$ and it can be expressed as
\begin{equation}
    h^2 \Omega_{\rm GW} \simeq h^2 \Omega_{\rm sw} + h^2 \Omega_{\rm col} + h^2 \Omega_{\rm turb}.
\end{equation}
\bl{For the EWPTs considered in this work we focus on the GW signal sourced by sound waves in the plasma. This approximation is appropriate for FOEWPTs proceeding in a thermal plasma in a non-runaway and plasma-dominated regime, where friction with the surrounding medium prevents indefinite acceleration of the bubble walls and most of the released energy is transferred to bulk fluid motion. In this regime, the long-lasting acoustic waves generated after bubble expansion and collision are expected to provide the dominant contribution to the GW spectrum \cite{Hindmarsh:2013xza, Hindmarsh:2015qta}. By contrast, the (pseudo)scalar field contribution from bubble-wall collisions is relevant mainly for runaway or strongly supercooled transitions, while the magnetohydrodynamic contribution from turbulence and magnetic fields is usually treated as a subleading and more uncertain source, generated by a fraction of the bulk kinetic energy of the plasma \cite{Caprini:2009yp}. The viable EWPTs found in our analysis are thermally driven rather than vacuum dominated and we therefore neglect the bubble-collision contribution and treat the magnetohydrodynamic turbulence contribution as indeed subleading. Furthermore, for simplicity, we fix the bubble-wall velocity to $v_w=1$, corresponding to the ultra-relativistic limit. This choice allows us to estimate the maximal sound-wave signal within our setup. We note, however, that $v_w$ can be determined dynamically and is generally model-dependent: dedicated computations in specific models often find values appreciably smaller than 1 \cite{Branchina:2025adj, Branchina:2025jou, DeCurtis:2024hvh, DeCurtis:2023hil, DeCurtis:2022hlx}. Therefore, our GW spectra should be interpreted as tentative estimates obtained in the ultra-relativistic wall-velocity limit.
}

For sound-waves, the state-of-the-art method to describe the spectral shape of the stochastic GW background produced by cosmological sources is through a Double Broken Power Law (DBPL), as shown in Ref.~\cite{Caprini:2024hue}. This parametrisation captures spectra whose slope changes at two characteristic frequencies. (In Appendix \ref{sec:BPL} we present a comparison with the GW spectra obtained by using the Broken Power Law (BPL) method~\cite{Carena:2025flp}). 
 
In the DBPL framework, the GW energy density spectrum can be written as
\begin{equation}
    \Omega^{\text{DBPL}}_{\text{GW}}(f)=\Omega_{\text{int}} \ S(f)=\Omega_{2} \  S_2(f),
\end{equation}
where $S(f)$ represents the dimensionless shape function of the spectrum. The parameter $\Omega_{\text{int}}$ corresponds to the integrated amplitude, while $\Omega_2$ denotes the amplitude evaluated at the second break frequency $\tilde{f}_2$. The shape function $S(f)$ is expressed as
\begin{equation}
    S(f)= N \left( \frac{f}{\tilde f_1} \right)^{n_{1}} \left[ 1 + \left(\frac{f}{\tilde f_1}\right)^{a_1}\right]^{- \frac{n_1-n_2}{a_1}}  \left[ 1 + \left(\frac{f}{\tilde f_2}\right)^{a_2}\right]^{- \frac{n_2-n_3}{a_2}},
\end{equation}
where $\tilde f_1$ and $\tilde f_2$ denote the two break frequencies. The spectral slopes are given by $n_1$, $n_2$ and $n_3$, describing the low, intermediate and high-frequency behaviour, respectively. The parameters $a_1$ and $a_2$ control how gradually the spectrum transitions between different power-law regimes around the two break frequencies. Regarding the  normalisation constant $N$, the spectrum is normalised at the second break frequency, defining
\begin{equation}
    S_2(f)=\frac{S(f)}{S(\tilde f_2)} \,.
\end{equation}
The two break frequencies are related to the characteristic scales of the fluid shell and the bubble size \cite{Caprini:2024gyk, Jinno:2022mie}, i.e.,
\begin{equation}
    \tilde f_1 \simeq 0.2 H_{*0} (H R_*)^{-1},  \ \ \ \ \ \ \ \ \ \ \ \ \ \ \ \ 
    \tilde f_2 \simeq 0.5 H_{*0} \Delta_w^{-1}(H R_*)^{-1},
\end{equation}
where $\Delta_w$ is the dimensionless shell thickness defined as $\Delta_w = 1 - c_s/v_w$ in the $v_w \to ~1$ limit, with $c_s= 1/\sqrt{3}$ being the speed of sound. The parameter $H_{*0}$  is the Hubble rate at the time of GW production, given by
\begin{equation}
H_{*0} = 1.65\times10^{-5} \ \text{Hz} \ \left(\frac{g_*}{100}\right)^{1/6}\left( \frac{T_n}{100 \  \text{GeV}}\right).
\end{equation}
The integrated amplitude of the spectrum can be written as \cite{Caprini:2024gyk, Jinno:2022mie}
\begin{equation}
    \Omega_{\text{int}}= F_{\text{GW,0}} A_{\text{SW}} K_{\text{SW}}^2 \Upsilon  (H R_*),
\end{equation}
with $\Upsilon = 1 - 1/ \sqrt{1 + 2 H \tau_{sw}}$ and $ H \tau_{\text{sw}} = \text{min}[1, 2 HR_*/\sqrt{ 3 K_{\text{sw}}}]$.

The bubble size $H R_*$ can be estimated as \cite{Caprini:2024hue}
\begin{equation}
(H R_*)= (8 \pi)^{1/3}\, v_w \,\frac{H}{\beta}
\end{equation}
and $F_{\text{GW},0}$ is the red-shift factor for the fractional energy density, i.e.,
\begin{equation}
h^2F_{\text{GW},0} = 1.64\times 10^{-5} \left(\frac{100}{g_*}\right)^{1/3}.
\end{equation}
The constant $A_{\text{SW}}\approx 0.11$ is obtained from numerical simulations. The parameter 
$K_{\text{SW}}$ represents the fraction of the energy density converted into kinetic energy of the plasma and can be approximated as
\begin{equation}
    K_{\text{SW}} \simeq \frac{0.6 \,\alpha\,\kappa_{\text{SW}}}{1 + \alpha},
\end{equation}
Here, $\kappa_{\text{SW}}$ describes the efficiency of converting the energy released during the transition into bulk motion of the plasma \cite{Espinosa:2010hh}, given by
\begin{equation}
    \kappa_{\text{SW}} \simeq \frac{\alpha}{0.73 + 0.083 \sqrt{\alpha}+ \alpha},
    \label{eq:kcoeff}
\end{equation}
\bl{which was obtained in the $v_w \to 1$ limit. (See also \cite{Giese:2020rtr,Giese:2020znk} for the estimates of $\kappa_{\text{SW}}$ beyond the bag model.)}
For the spectral slopes of the DBPL template describing sound waves, numerical results \cite{Jinno:2022mie} indicate the values
\begin{equation}
    n_1=3, \ \ \ \ 
    n_2=1, \ \ \ \ 
    n_3=-3, \ \ \ \
    a_1=2, \ \ \ \
    a_2 = 4.
\end{equation}
With this choice, the amplitude at the second break frequency is related to the integrated amplitude via
\begin{equation}
    \Omega_2 = \frac{1}{\pi}\left( \sqrt{2} +\frac{2 \tilde f_2/ \tilde f_1}{1+ \tilde f^2_2/\tilde f^2_1}\right)\ \Omega_{\text{int}}.
\end{equation}
{We note that the GW spectrum is actually controlled by the percolation temperature. However, as mentioned in the previous section, in the A2HDM, they are essentially the same.} 

\subsection{Implications for LISA}
To assess the observational prospects, we compare the predicted GW spectra with LISA's sensitivity. The detectability of the signal is quantified through the SNR quantity \cite{Babak:2021mhe}, i.e., 
\begin{equation}
\mathrm{SNR} =
\left[
t_{\rm obs}
\int_{f_{\rm min}}^{f_{\rm max}}
\mathrm{d}f\,
\left(
\frac{h^2\Omega_{\rm GW}(f)}
     {h^2\Omega_{\rm Sens}(f)}
\right)^2
\right]^{1/2},
\end{equation}
 where $t_{\rm obs}$ is the observation time and $h^2\Omega_{\rm GW}(f)$ is the predicted GW energy-density spectrum. 
 
 The sensitivity curve $\Omega_{\rm Sens}(f)$ is related to the strain noise
power spectral density $S_h(f)$ according to
\begin{equation}
\Omega_{\rm Sens}(f) =
\frac{2\pi^2}{3H_0^2}\, f^3 S_h(f).
\end{equation}
Here, the present-day value of the Hubble parameter is assumed to be
$H_0 = 2.19 \times 10^{-18}\,\mathrm{s}^{-1}$. This conversion allows us to express the detector noise in terms of a dimensionless energy-density spectrum, directly comparable to the predicted stochastic GW signal.

For LISA, the strain noise power spectral density is modelled as
\begin{equation}
S_h(f) =
\frac{20}{3}
\left[
\frac{5.76 \times 10^{-48}}{(2\pi f)^4}
\left(1 + \frac{f_1^2}{f^2}\right)
+ 3.6 \times 10^{-41}
\right]
\left(1 + \frac{f^2}{f_2^2}\right),
\end{equation}
where $f_1 = 0.4 \times 10^{-3}\,\mathrm{Hz}$ and $f_2 = 25 \times 10^{-3}\,\mathrm{Hz}$ 
define the low and high frequency transitions, respectively \cite{Babak:2021mhe}. 

In our analysis, we consider the observation time 
$t_{\text{obs}} = 3\,\mathrm{yrs}$, and the integration is performed over the frequency range $[f_{\text{min}}, f_{\text{max}}]$ accessible to LISA, i.e., $f_{\rm min} = 10^{-6}\,\mathrm{Hz}$ to $f_{\rm max} = 10\,\mathrm{Hz}$, when computing the SNR.

\subsection{Interpretation of Finite Temperature Results}

\begin{figure}[h!]
    \centering
    \includegraphics[width=0.8\textwidth]{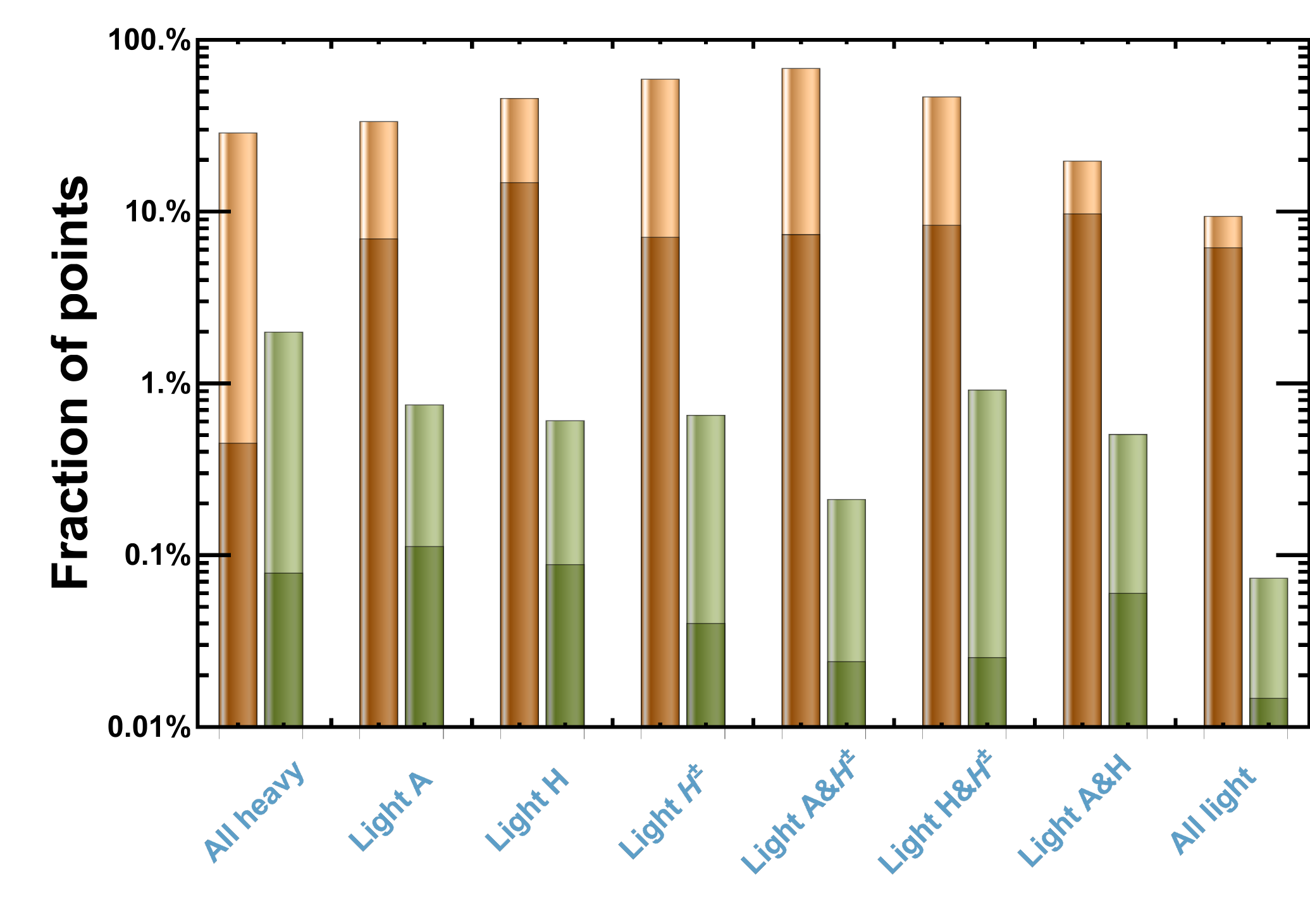}
    \caption{Brown bars: Fraction of allowed points from {\tt HEPfit} generating FOEWPTs.
    Green bars: Fraction of points from {\tt HEPfit} capable of generating \snr{} during GW detection at LISA. In both the bars, darker colors represent multi-step FOEWPTs.}
    \label{fig:1st_order_frac}
\end{figure}
While the spectral shape remains largely universal, as in other BSM extensions, differences arise at the quantitative level due to the distinct regions of parameter space allowed by theoretical and experimental constraints. This leads to variations in the EWPT parameters compared to other 2HDM realisations resulting in differences in the distributions of peak frequency and SNR of the GW signal~\cite{Goncalves:2021egx,Lee:2025hgb,Wang:2019pet,Benincasa:2022elt,Arcadi:2022lpp}.

Since the amplitude and the peak frequency are directly controlled by $\alpha$ and $\beta/H$, it is useful to examine the behaviour of these parameters across the scenarios considered. In this section, we present all the results we have obtained, providing a comprehensive overview of the parameter space and its implications for the GW signatures. To investigate the EWPT and the associated GW signals, we generate 75,000 distinct points in each of the eight regions of the parameter space (see Sec.\ref{sec:parameters}) using posterior distribution of {\tt HEPfit}. These points are subsequently analysed with {\tt CosmoTransitions}, and we retain those exhibiting a FOEWPT, independently of the number of steps, and complying with EW symmetry restoration at high temperature. For the remaining viable points, we compute the peak frequency and amplitude of the resulting GW signals as observed today, and evaluate their detectability at LISA by estimating the SNR.

\vspace*{3mm}

In Fig. \ref{fig:1st_order_frac}, we show the efficiencies of all eight regions of the parameter space in terms of the occurrence of FOEWPTs and the corresponding SNR at LISA. The brown bars indicate the fraction of potential points yielding a strong FOEWPT, following the selection criteria outlined above, while the green bars denote the fraction of points producing \snr{} at LISA. In both cases, the darker shading highlights the subset of multi-step transitions while the lighter one represents the subset of one-step transitions, with the full bar indicating the total fraction of points that exhibit FOEWPT.

{It is evident that the region of parameter space with both $A$ and $H^\pm$ light exhibits the highest fraction of points ($\sim70\%$) undergoing a strong FOEWPT, however, very few of these points (only $0.2\%$) generate a large SNR at LISA. Conversely, in the region where all BSM Higgs states are heavy, approximately 30\% of points lead to a strong FOEWPT, with 2\% of them achieving \snr, the highest among all scenarios. Further, a light $H$ leads to the largest number of multi-step transitions, while the case with all scalars being heavy results in the fewest. 
This is consistent with expectations: in a light 
$H$ scenario  a tree‑level barrier in the $(h,H)$ field space can be generated, favouring two‑step transitions, whereas for a heavy 
$H$
 the transition typically occurs directly along the 
$h$-direction through a one‑loop–induced barrier. \bl{Among multistep transitions, the most frequent are two-step transitions, following the pattern $(0,0) \to (h_1,H_1) \to (h_2,H_2)$. In the following analysis, we consider all first-order transition steps, regardless of whether they occur as part of a one-step or multistep transition sequence.}

At any rate, in most scenarios, one-step transitions are more prevalent than multi-step transitions. Exceptions occur in the region with both $A$ and $H$ light, where number of one-step and multi-step transitions are comparable, as well as in the scenario with all scalars light, where multi-step transitions dominate. 

Regarding the SNR values, the region with all scalars light performs most poorly, with only $\sim0.07\%$ of points exceeding \snr.}  Notice also that the SNR values can be very large, above and beyond what seen in the standard Yukawa types of the 2HDM, owing primarily to the significantly larger parameter space pertaining to the A2HDM. However, as we shall see in Appendix \ref{sec:BPL}, they represent a statistically insignificant ensemble (even for the most favourable scenario mentioned above where all BSM Higgs states are heavy).

\begin{figure}[h!]
\centering
\subfloat[\textbf{All Heavy}]{%
    \includegraphics[scale=0.28]{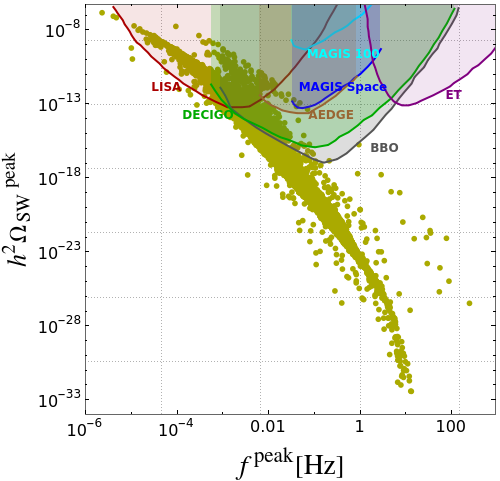}%
}%
\hspace*{0.15truecm}
\subfloat[\textbf{ Mixed Heavy \& Light}]{%
    \includegraphics[scale=0.28]{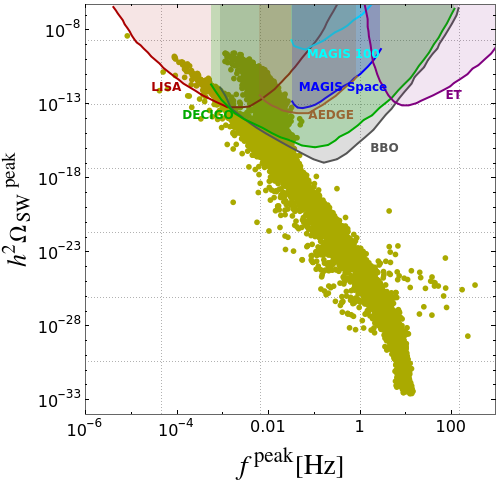}%
}%
\hspace*{0.15truecm}
\subfloat[\textbf{All Light}]{%
    \includegraphics[scale=0.28]{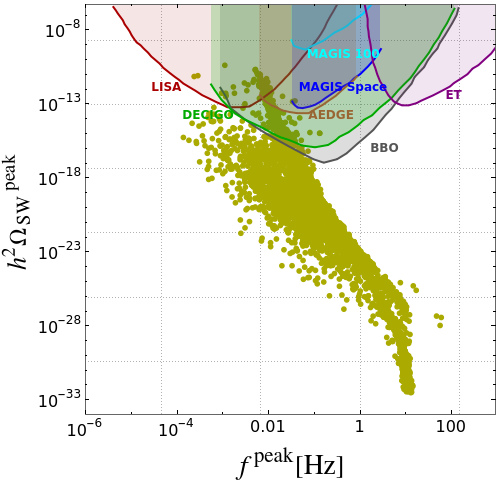}%
}
\caption{GW spectra for the different A2HDM scenarios. 
Panels (a), (b) and (c) correspond to the ``All Heavy'', ``\text{Mixed Heavy \& Light}'' and ``All Light'' 
mass configurations, respectively. The green points represent the predicted peak amplitudes 
$h^2 \Omega_{\rm GW}^{\rm peak}$ and peak frequencies $f_{\rm peak}$ for each scenario. 
The coloured shaded regions indicate the projected sensitivities of current and future 
GW detectors. }
\label{fig:GWspectra}
\end{figure}

\noindent
$\bullet$ $\bm{h^2\Omega_{SW}^{peak}}$ vs $\bm {f^{peak}}$

{Fig.~\ref{fig:GWspectra} shows the distribution of the peak frequency and the corresponding GW amplitude for all the eight regions of the A2HDM parameter space. The left panel corresponds to the scenario in which all scalars are heavy, the right panel shows the scenario with all scalars light and the middle panel combines the remaining six scenarios with proper weights featuring mixed heavy and light scalars. We find that none of the explored regions of parameter space (and hence A2HDM) are detectable by MAGIS-100 \cite{Graham:2016plp, Graham:2017pmn} or the Einstein Telescope (ET) \cite{Punturo:2010zz, Hild:2010id}. In contrast, all eight scenarios yield signals that are potentially detectable by other interferometers, such as LISA \cite{LISA:2017pwj, Baker:2019nia}, DECIGO  \cite{Yagi:2011wg, Kawamura:2020pcg}, BBO \cite{Crowder:2005nr, Corbin:2005ny}, AEDGE \cite{AEDGE:2019nxb, Badurina:2021rgt} and {\black MAGIS} \cite{Graham:2016plp, Graham:2017pmn}, although a sizeable fraction of the parameter points lie below the corresponding sensitivity curves. In particular, the scenario with all BSM Higgses  heavy can be readily detected by LISA, even at very low peak frequencies of order $10^{-5}$ Hz. The mixed scenarios are also easily detectable at LISA, but typically at comparatively higher peak frequencies. By contrast, the scenario with all (pseudo)scalars light is only marginally detectable at LISA.

\begin{figure}[!t]
\centering
\includegraphics[width=\linewidth]{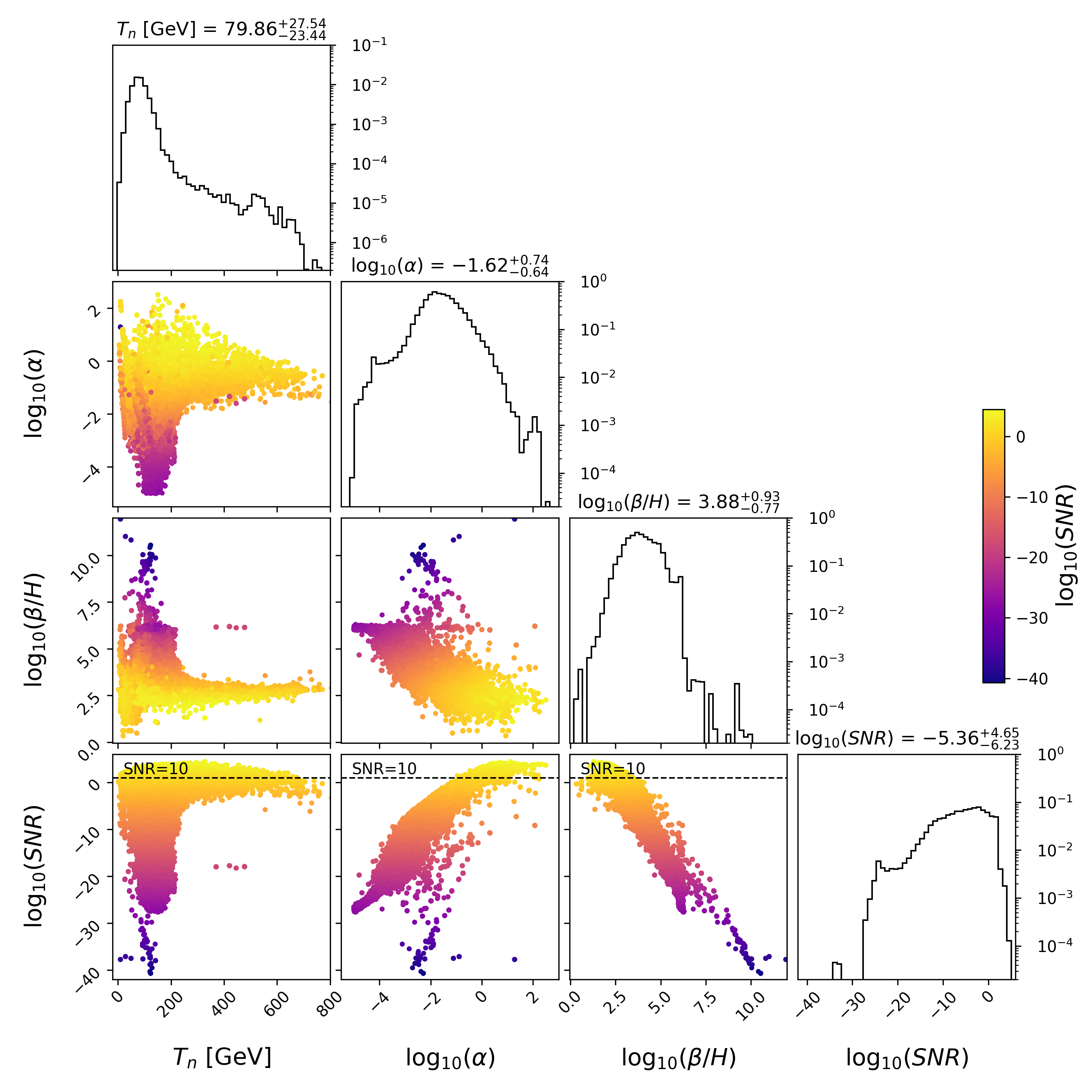}
\caption{Combined correlation plots between parameters of the A2HDM over all eight regions discussed. The variables shown are the nucleation temperature $T_n$; the logarithm of the transition strength, $\log_{10}(\alpha)$; of the inverse duration, $\log_{10}(\beta/H)$; 
and of the SNR, $\log_{10}({\rm SNR})$. Each off-diagonal panel shows a scatter plot of two parameters, with points coloured by $\log_{10}({\rm SNR})$, where yellow indicates higher SNR values and dark purple lower SNR. Diagonal panels display the one-dimensional probability distribution of each parameter along with the central value and 68\% probability region around it.}
\label{fig:CornerPlot}
\end{figure}

\vspace*{3mm}
\noindent
$\bullet$ \textbf{Correlations} 

{In Fig.~\ref{fig:CornerPlot}, we present the correlations among the key parameters of the A2HDM that give rise to strong FOEWPTs\footnote{For the behaviour of $v_c/T_c$ see Appendix \ref{sec:vcbytc}.}.
To this end, we combine all eight regions of the parameter space using appropriate weights. The diagonal panels display the one-dimensional marginalised distributions, while the off-diagonal panels show the projected two-dimensional correlations, colour coded by the SNR.}
While the two-dimensional correlation plots indicate that a non-negligible population of parameter points, primarily associated with mixed heavy and light scenarios, exhibits elevated $T_n$, the one-dimensional histogram of $T_n$ reveals that  the majority of the sampled parameter space is characterised by $T_n \lesssim 200$ GeV.

Concerning the GW parameters, the peaking behaviour in the $(\beta/H-T_n)$ plane is correlated to an inverse relationship observed in the $\alpha$ and SNR  distributions relative to $T_n$ which exhibit characteristic troughs in the vicinity of EW scale. As such, it is sufficient to comment the behaviour of $\beta/H$ on $T_n$.
A pronounced clustering of points is observed in the range $T_n \in (100 \GeV,\,160\GeV)$, accompanied by a significant spread in $\beta/H$. This indicates that, even within a relatively narrow temperature interval, the dynamics of the FOEWPT can vary substantially. 
The peak in $\beta/H$ as a function of $T_n$ can be understood on general grounds, since it occurs near the EW scale, equivalently close to the critical temperature. When $T_n \simeq T_c$, the amount of supercooling is small and the phase transition is therefore expected to be weak and of short duration, resulting in a large value of $\beta/H$.
The accumulation of points at large $\beta/H$ is associated with the large number of weak transitions, in agreement with Fig.~\ref{fig:1st_order_frac}. \bl{We have also explicitly checked that, for all points with low nucleation temperatures (e.g. $T_n < 30,\mathrm{GeV}$), removing the thermal masses entering the high-temperature resummation prescription leads to negligible changes in $T_n$ and $\beta/H$, and to variations of at most $\sim 10\%$ in $\alpha$, leaving our conclusions unaffected.}

Across all scenarios, very large values of $\beta/H$ are present over a wide portion of the parameter space. These correspond to very rapid EWPTs, which in turn produce GW spectra shifted toward higher frequencies and characterised by extremely low SNR, rendering them effectively undetectable at interferometers such as LISA.

While the one-dimensional histograms of the parameters $\alpha, \beta/H$ and the resulting SNRs suggest that the bulk of the parameter space yields signals below the LISA sensitivity threshold, the correlation plots provide a more nuanced perspective. We identify a significant ensemble of points defined by a large latent heat ($\alpha$) and suppressed inverse-time duration ($\beta/H$), i.e., the yellow points. These points correspond to an enhanced vacuum energy release, which sufficiently amplifies the stochastic GW  background to meet or exceed the LISA detection limits. As shown in Appendix~\ref{sec:correlation}, these points with large SNR are scattered around the parameters space of the A2HDM. 
Although the conventional 2HDM frameworks can provide large SNR at LISA (see for instance \cite{Goncalves:2023svb,Goncalves:2021egx,Lee:2025hgb}, in which the BPL formalism has been used), the parameter space of such frameworks are usually more restrictive than A2HDM.

\section{Collider Searches for A2HDM Signals}
\label{sec:collider}

In the previous section, we identified regions of the A2HDM parameter space that give rise to a strong FOEWPT and a sufficiently large SNR for GW detection. In this section, we investigate the collider phenomenology of these viable parameter points and assess their testability at the HL-LHC.

Our analysis focuses on the major production and decay channels of the additional neutral scalar and pseudo-scalar states of the model, $H, A$,
as well as the charged Higgs states, $H^{\pm}$, when all cross sections of the SM-like Higgs state, $h$, are close to their SM limits.
 For the parameter points selected from the EWPT analysis, we compute production cross sections and decay BRs for the relevant channels and compare these with projected sensitivities for the HL-LHC. The main goal here is to identify the region of A2HDM parameter space that can give rise to a strong FOEWPT, strong GW signals (SNR $> 10$) that may be detectable in the future (e.g., by LISA), and at the same time have the potential to be observed at the HL-LHC. In fact, we will also be looking at the case with SNR $> 30$, i.e., when diagnostic of a GW signal (in terms of the source and its properties)  may become possible.

\subsection{Signal Processes and Simulation Setup}
From a theoretical perspective, the parameter regions of the A2HDM that support a strong FOEWPT can exhibit sizeable scalar self-couplings and a variety of Higgs-mass spectra, ranging from compressed to more hierarchical configurations. Depending on the mass pattern and couplings, ggF production of heavy neutral (pseudo)scalar Higgs states can be enhanced, and bosonic decay modes such as \(H \to WW,\, ZZ,\, hh\), as well as cascade decays like \(A \to Zh\), can become important, as is typical of extended Higgs sectors. Fermionic decay channels, such as \(H \to \tau^+\tau^-\), \(H, A \to t \bar{t}\) and \(H^\pm \to \tau^\pm \nu_\tau\), also remain relevant due to their clean experimental signatures and low backgrounds.

Experimentally, the channels considered here correspond to those for which the LHC currently provides the most stringent constraints over a wide mass range. In fact, ggF production of non-standard neutral (pseudo)scalar Higgs states followed by decays into vector bosons, di-Higgs, or  $t \bar{t}$ associated production of $H, A$ with $4t$ final states are the primary sensitivity probes for high $H, A$ masses, while ggF production of $H$, decaying to ($\tau$) leptons are very powerful to probe low mass regions of additional (pseudo)scalar states as well. The \(pp \to tbH^\pm\) with \(H^\pm \to \tau^\pm \nu_\tau\) dominates the probes of charged Higgs bosons with mass above $\sim 80-90$ GeV (i.e., the LEP limit from charged Higgs boson direct searches \cite{ALEPH:2013htx}). Other production modes, such as Vector Boson Fusion (VBF), $VH$, $VA$, or $H^+H^-$, are generally subdominant and provide fewer competitive constraints in the relevant mass ranges highlighted here in the EWPT analysis.

Explicitly, 
the ggF mode,
\begin{equation}
gg \to H, A,
\end{equation}
proceeds through heavy (mainly top and marginally bottom) quark loops and is particularly sensitive to the Yukawa structure of the A2HDM, and so is the single $H^\pm$ channel.
\begin{equation}
pp \to t\bar  b H^+ +~{\rm c.c.}
\end{equation}
{\color{black}{For each viable parameter point, the inclusive production cross sections \(\sigma(gg \to H, A)\), \(\sigma(pp \to t \bar{t} H, A)\) and \(\sigma(pp \to t b H^{\pm})\), together with the corresponding decay BRs into SM final states, are computed at Leading Order (LO) using \texttt{MadGraph5\_aMC@NLO}~\cite{Alwall:2014hca,Frederix:2018nkq}. Throughout this analysis, we employ the \texttt{MMHT2014lo68cl} Parton Distribution Function (PDF) set as implemented in \texttt{LHAPDF6}~\cite{Buckley:2014ana}. The renormalisation and factorisation scales are identified, i.e.,  $\mu_R\equiv\mu_F$, and chosen according to the default dynamical-scale prescription of \texttt{MadGraph5\_aMC@NLO}. For the resonant heavy Higgs production processes considered here, this corresponds to a characteristic scale of the order of its mass. 
In fact, we have verified that the predicted production cross sections are only weakly dependent on the choice of PDF sets. Comparisons with alternative PDF parametrisations, including \texttt{MSHT20lo\_as130}, lead to variations well below 5\%, leaving all collider projections and conclusions of this analysis unchanged.}}

The decay patterns of heavy Higgs states are strongly model-dependent. In this study, we focus on the channels most relevant for collider searches:  
\begin{align}
H &\to WW, \quad ZZ, \quad hh, \quad \tau^+ \tau^-, \quad t \bar{t},\\
A &\to Zh, \\
H^{\pm} &\to \tau^{\pm} \nu_{\tau}.
\end{align}  
For neutral (pseudo)scalars, we consider  
\begin{equation}
\sigma(pp \to t\bar{t} H, A) \times \mathrm{BR}(H, A \to t \bar{t})
\end{equation}  
and
\begin{equation}
\sigma(gg \to H, A) \times \mathrm{BR}(H, A \to X_j X_k),
\end{equation} 
where \(X_j, X_k\) denote the final-state particles, while for the charged Higgs, we consider 
\begin{equation}
\sigma(pp\to t b H^{\pm}) \times \mathrm{BR}(H^{\pm} \to \tau^{\pm} \nu_{\tau}),
\end{equation}  
as a function of the (pseudo)scalar mass and compare these predictions with existing LHC limits and future projections.

\subsection{Current LHC Limits and Future HL-LHC Projections}
\label{subsec:LHC_ref}
Experimental searches for additional Higgs bosons at the LHC are typically
reported as upper bounds on the production rate,
$\sigma \times \mathrm{BR}$, at a CoM energy of
$\sqrt{s} = 13~\mathrm{TeV}$ and for a given integrated luminosity. A few current references are given below, which have been used in this paper as illustrative examples of existing LHC limits.

\begin{itemize}
  \item ATLAS has set 95\%~CL limits on $\sigma(gg \to A)\times {\rm BR}(A\to Zh)$ 
  from the $A\to Zh$ decay into leptons and $b$ jets at $\sqrt{s}=13$~TeV~\cite{ATLAS:2022enb}. CMS also provides a 95\%~CL limit on the same in ~\cite{CMS:2025bvl}.
  \item ATLAS~\cite{ATLAS:2024itc, ATLAS:2024zoq, ATLAS:2020tlo} has published limits on heavy resonances decaying to vector boson pairs as well as di-Higgs pairs, whereas CMS~\cite{CMS:2018kaz} has also published limits on heavy resonances decaying to Higgs pairs ($hh$) at $\sqrt{s}=13$~TeV.
  \item CMS has also constrained resonant Higgs pair production in channels involving $WW^*WW^*$, $WW^*\tau\tau$ and $\tau\tau\tau\tau$ final states~\cite{CMS:2021roc}.
  \item CMS heavy Higgs searches in the $\tau^+\tau^-$ channel at 13~TeV set limits on $\sigma\times {\rm BR}(H,A\to\tau\tau)$~\cite{CMS:2018rmh}.
  \item ATLAS has also published limits on $H,A \to t \bar{t}$, via ggF production ~\cite{ATLAS:2024itc, ATLAS:2024vxm} and $t \bar{t} H,A$ ~\cite{ATLAS:2022rws} production channels at $\sqrt{s}=13$~TeV.
  \item Charged Higgs boson searches by ATLAS constrain $\sigma(pp\to tbH^\pm)\times {\rm BR}(H^\pm\to\tau\nu)$ over wide mass ranges~\cite{ATLAS:2024itc}.
\end{itemize}

To assess the sensitivity of the HL-LHC,
which is expected to operate at $\sqrt{s} = 14~\mathrm{TeV}$ with an
integrated luminosity of $\mathcal{L} = 3~\mathrm{ab}^{-1}$, it is
necessary to extrapolate these limits to the new collider configuration.

The improvement in sensitivity arises from two distinct effects: the
increase in integrated luminosity and the change in the proton-proton
CoM energy. While the former can be accounted for by a simple
statistical rescaling, the latter affects the production cross sections in
a process-dependent manner through the partonic structure of the proton.
In the following, we describe the procedure adopted to perform this
extrapolation for both neutral and charged Higgs bosons in the A2HDM.

Throughout this section, we assume that detector acceptances, selection
efficiencies and systematic uncertainties remain approximately unchanged
between 13 and 14~TeV. Under this assumption, the extrapolation of the
experimental limits is governed entirely by the scaling of the signal
cross section, $\sigma$, and the integrated luminosity, ${\cal L}$.

\subsection{Scaling of $gg\to H,A$ Production}
\label{subsec:ggf_scaling}

As mentioned, the dominant production mechanism for the neutral Higgs bosons
$H, A$ over a wide mass range at the LHC is ggF production.
Within the QCD factorisation framework, the inclusive production cross
section at a proton-proton collider with CoM, $\sqrt{s}$
can be written as
\begin{equation}
\sigma_{gg \to H, A}(s)
=
\int_0^1 dx_1 \, dx_2 \;
g(x_1,\mu)\,
g(x_2,\mu)\,
\hat{\sigma}_{gg \to H, A}(\hat{s},\mu),
\end{equation}
where $g(x,\mu)$ denotes the gluon PDF, $\mu$ is the factorisation/renormalisation scale and
$\hat{s} = x_1 x_2 s$ is the partonic CoM energy squared.

For an on-shell (pseudo)scalar with mass $m_{H, A}$ and a total width much smaller
than its mass, the narrow-width approximation applies. In this limit, the
partonic cross section is effectively localised at
$\hat{s} = m_{H, A}^2$ and the hadronic cross section can be expressed in
terms of the gluon-gluon parton luminosity,
\begin{equation}
\sigma_{gg \to H, A}(s)
=
\frac{1}{s}
\left.
\frac{d\mathcal{L}_{gg}}{d\tau}
\right|_{\tau = m_{H, A}^2/s}
\,
\hat{\sigma}_{gg \to H, A}(\hat{s}=m_{H, A}^2),
\end{equation}
with
\begin{equation}
\tau \equiv \frac{m_{H, A}^2}{s},
\qquad
\frac{d\mathcal{L}_{gg}}{d\tau}
=
\int_\tau^1 \frac{dx}{x}\,
g(x,\mu_F)\,
g\!\left(\frac{\tau}{x},\mu_F\right).
\end{equation}

The quantity $\hat{\sigma}_{gg \to H, A}$ encodes the model-dependent
couplings and loop structure of the Higgs boson and therefore differs for the
CP-even and CP-odd states of the A2HDM. However, once evaluated at
$\hat{s} = m_{H, A}^2$, it does not depend explicitly on the hadronic
CoM energy. As a result, the entire $\sqrt{s}$ dependence of
the hadronic cross section is governed by the gluon PDFs.

This observation allows one to relate ggF production cross sections at
different collider energies in a model-independent way. For two
CoM energies $\sqrt{s_1}$ and $\sqrt{s_2}$, the ratio of cross
sections is given by
\begin{equation}
\frac{\sigma_{gg \to H, A}(s_2)}{\sigma_{gg \to H, A}(s_1)}
=
\frac{
\left.
\dfrac{d\mathcal{L}_{gg}}{d\tau}
\right|_{\tau = m_{H, A}^2/s_2}
}{
\left.
\dfrac{d\mathcal{L}_{gg}}{d\tau}
\right|_{\tau = m_{H, A}^2/s_1}
}.
\label{eq:gg_scaling_hi}
\end{equation}

In this work, Eq.~\eqref{eq:gg_scaling_hi} is used to extrapolate the
existing 13~TeV limits on $\sigma\times\mathrm{BR}$ to
14~TeV. The gluon-gluon luminosities are evaluated numerically using the
same PDF set and scale choices. Although the
relation above is formally derived at LO, higher-order QCD
corrections to ggF largely factorise, such that the ratio in
Eq.~\eqref{eq:gg_scaling_hi} is only weakly dependent on the perturbative
order when consistent inputs are used.

\subsection{Scaling of $gg,q\bar q\to t\bar{t}H,A$ and $tbH^{\pm}$ Production}
\label{subsec:tbh_scaling}

For charged Higgs bosons, the dominant production mechanism at the LHC for
$m_{H^\pm} \gtrsim m_t$ is associated production with a top and bottom
quark, $pp \to tbH^\pm$, followed in our analysis by the decay
$H^\pm \to \tau^\pm \nu_\tau$. Also there is another major bound that can come from $t \bar{t}H,A$ production and decay of $H,A$ into top pairs. Unlike neutral Higgs production via ggF, this process receives contributions from multiple initial states and
kinematic configurations, so that its energy dependence cannot be expressed
in terms of a single partonic luminosity.

Dedicated studies of charged Higgs production in 2HDMs,
nevertheless, show that the ratio
\begin{equation}
R^{H^\pm}_{14/13}
=
\frac{\sigma(pp \to tbH^\pm,~14~\mathrm{TeV})}
{\sigma(pp \to tbH^\pm,~13~\mathrm{TeV})}
\end{equation}
varies smoothly with the charged Higgs mass, increasing from values close
to unity at low masses to approximately $1.3$--$1.4$ in the several
hundred GeV range. This behaviour is driven primarily by the change in the
relevant parton momentum fractions and the increased phase space at
14~TeV. Similarly, for $t\bar{t}H,A$, the scaling factor varies from $1.1$ to $1.3$ in the mass range of 200 GeV to 1 TeV.

Given the phenomenological nature of the HL-LHC projection and the fact
that the dominant improvement arises from the substantial increase in
integrated luminosity, we adopt two representative constant values, 
\begin{equation}
R^{H^\pm}_{14/13} = 1.3
\end{equation}
and
\begin{equation}
R^{(H,A)_{t\bar{t}H,A}}_{14/13} = 1.2,
\end{equation}
to rescale the current 13~TeV limits for these two analyses. This choice is intended for qualitative projections rather than precision forecasting, while nonetheless providing a reasonable estimate of the expected sensitivity over the
mass range considered and not affecting the conclusions of
our analysis. (For example, we have verified that varying this factor within the range
$1.2$--$1.4$ has a negligible impact on the projected reach).

\subsection{Inclusion of Luminosity Effects}
\label{subsec:lumi_scaling}

Once the cross section limit has been rescaled to $\sqrt{s}=14~\mathrm{TeV}$, the effect of the increased integrated luminosity is incorporated. Assuming that the analyses remain dominated by statistical uncertainties and that detector acceptance and signal efficiencies do not change significantly between Run 3 of the LHC and the HL-LHC, the expected upper limit on the cross section scales approximately as
\begin{equation}
\left(\sigma \times \mathrm{BR}\right)_{14}^{\mathrm{HL}}
=
\left(\sigma \times \mathrm{BR}\right)_{14}
\times
\sqrt{\frac{\mathcal{L}_{13}}{\mathcal{L}_{14}}},
\label{eq:lumi_scaling}
\end{equation}
where $\mathcal{L}_{13}$ denotes the integrated luminosity of the current LHC dataset and $\mathcal{L}_{14}$ is the target luminosity.

In this work, we take $\mathcal{L}_{13} = 139~\mathrm{fb}^{-1}$ for $H \to W^+ W^-$, $H \to ZZ$, $H \to hh$, $A \to Zh$ for ggF production of $H$ and $A$, $\mathcal{L}_{13} = 35.9~\mathrm{fb}^{-1}$ for ggF production of $H \to \tau^+ \tau^-$, $\mathcal{L}_{13} = 139~\mathrm{fb}^{-1}$ for $t\bar{t}H,A$ production followed by $H,A \to t \bar{t}$ and $\mathcal{L}_{13} = 36.1~\mathrm{fb}^{-1}$ for $tbH^{\pm}$ production followed by $H^{\pm} \to \tau^{\pm} \nu_{\tau}$. Then, $\mathcal{L}_{14} = 3~\mathrm{ab}^{-1}$ has been used for HL-LHC scaling. The combined energy and luminosity rescaling procedure yields projected HL-LHC sensitivity curves that can be directly compared with the predictions of the A2HDM parameter points identified in the previous sections. This allows us to assess the collider reach of the HL-LHC in regions of parameter space associated with a strong FOEWPT and potentially observable GW signals.

\begin{figure}[htpb!]
    \centering
    \includegraphics[width=0.49\textwidth]{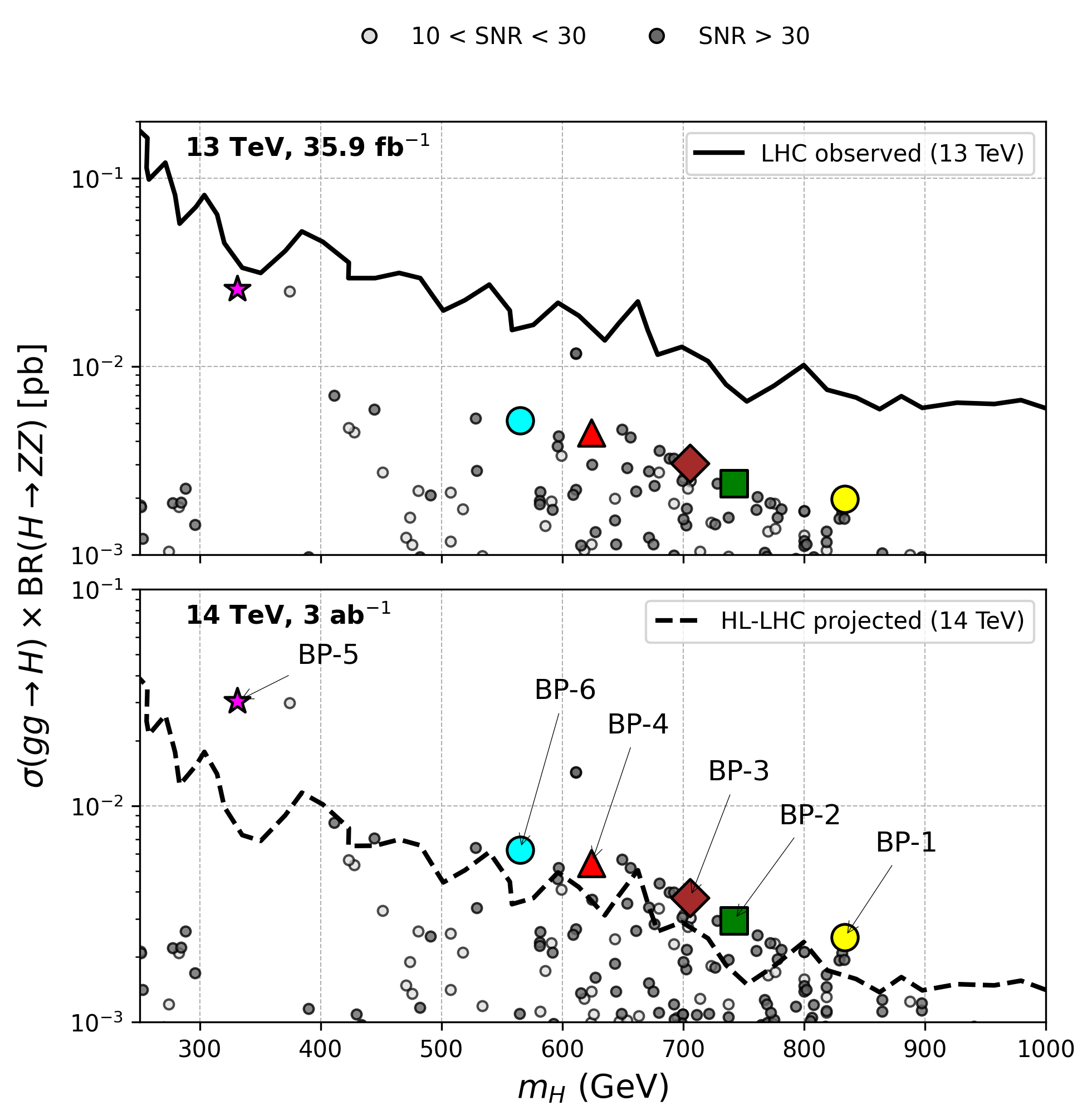}
    \includegraphics[width=0.49\textwidth]{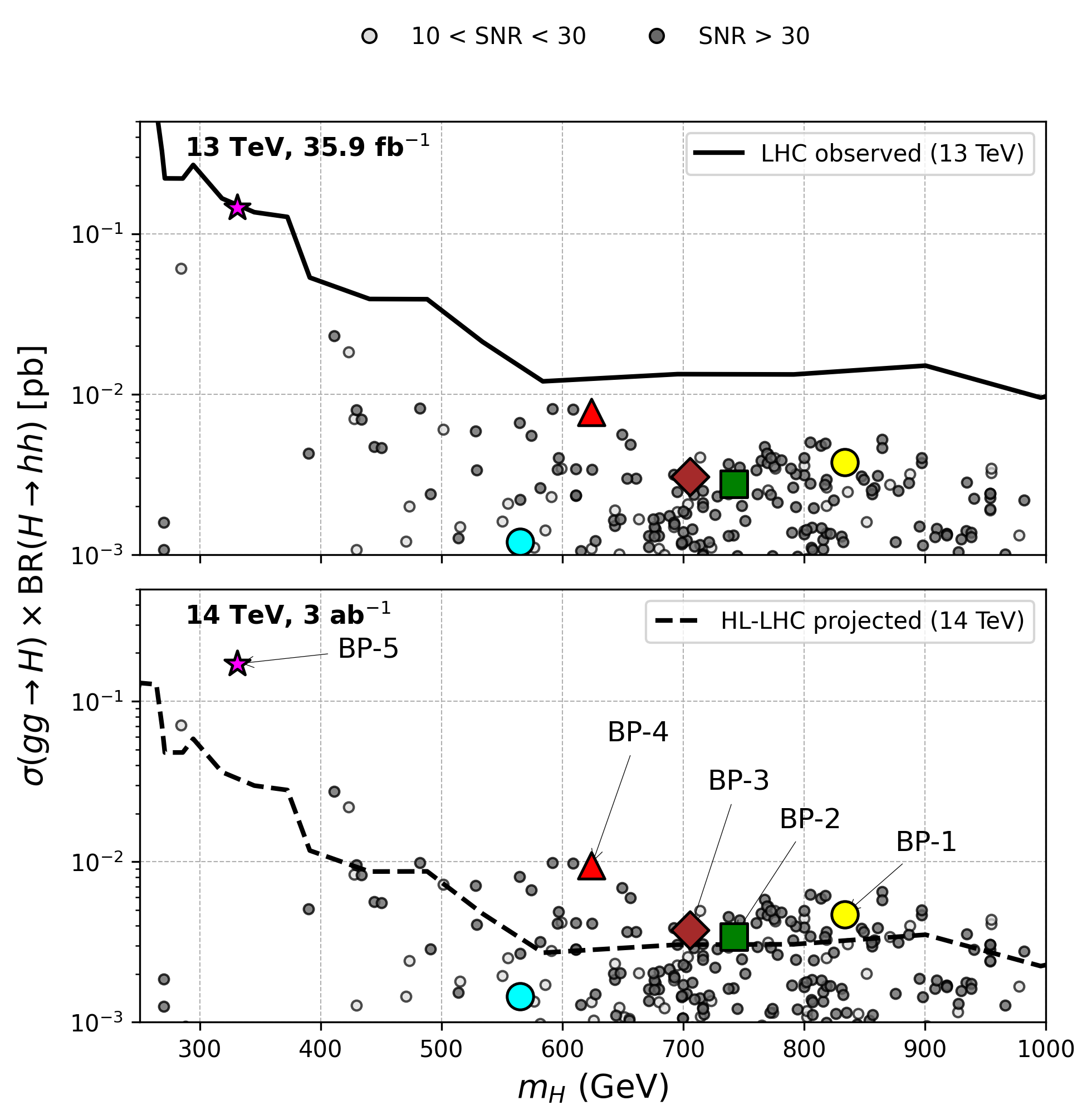}
    \caption{Predicted ggF production rates times decay BRs in the $H \to ZZ$ (left) and $H \to hh$ (right) channels as a function of $m_H$ for the 13 TeV LHC (top) and 14 TeV LHC (bottom) for parameter points consistent with a strong FOEWPT and an observable GW signal. Light gray and dark gray points represent the 10 < {\rm SNR} < 30 and SNR > 30 regions, respectively. The solid black line denotes the current Run 3 upper limit, while the dashed black line shows the projected HL-LHC sensitivity 
obtained by rescaling such a limit using gluon-gluon parton luminosities and
an integrated luminosity of $3~\mathrm{ab}^{-1}$. Also shown are some of the Benchmark Points (BPs) defined in Table~\ref{tab:bp_summary}. }
\label{fig:Collider_HZZ_Hhh}
\end{figure}

\begin{figure}[h!]
    \centering
    \includegraphics[width=0.49\textwidth]{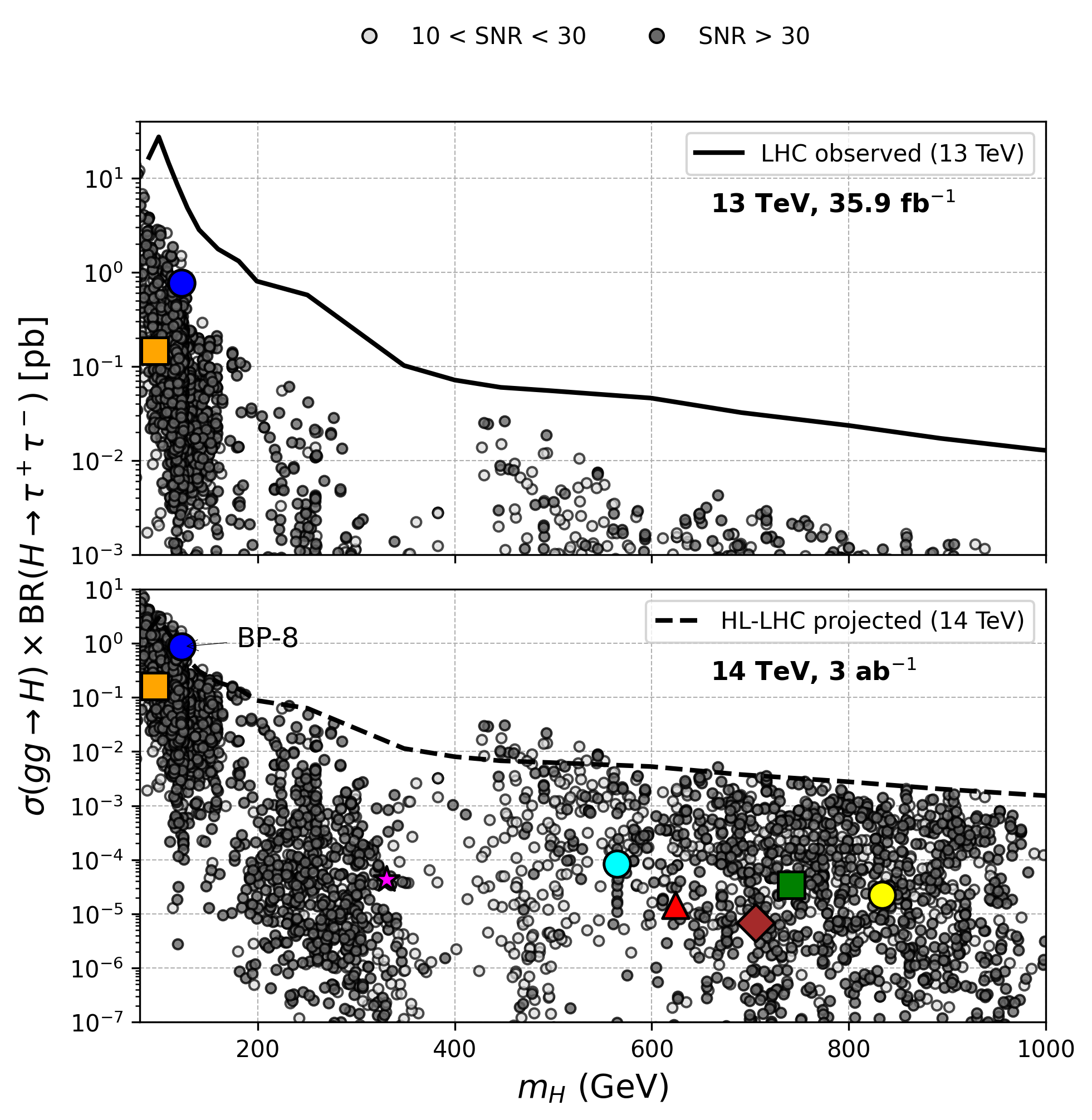}
    \includegraphics[width=0.49\textwidth]{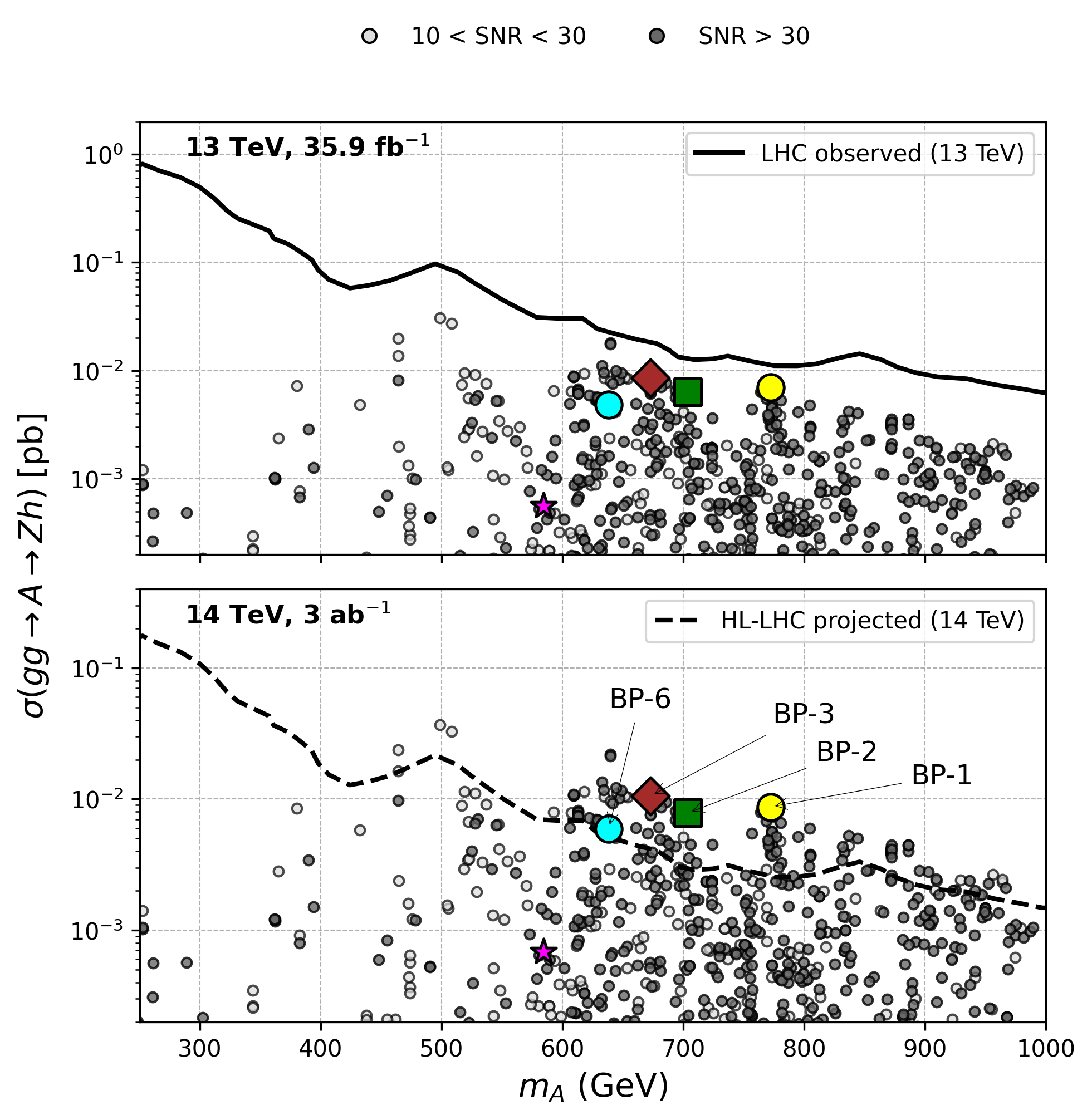}
    \caption{Same as Fig.~\ref{fig:Collider_HZZ_Hhh}, but for the $H \to \tau^+ \tau^-$ (left) and  $A \to Zh$ (right) channels. The gaps in the distribution of points are related to having plotted only configurations with SNR $>10$ (see Appendix~\ref{sec:correlation}).}
    \label{fig:Collider_Htautau_AZh}
\end{figure}

\begin{figure}[h!]
    \centering
    \includegraphics[width=0.6\textwidth]{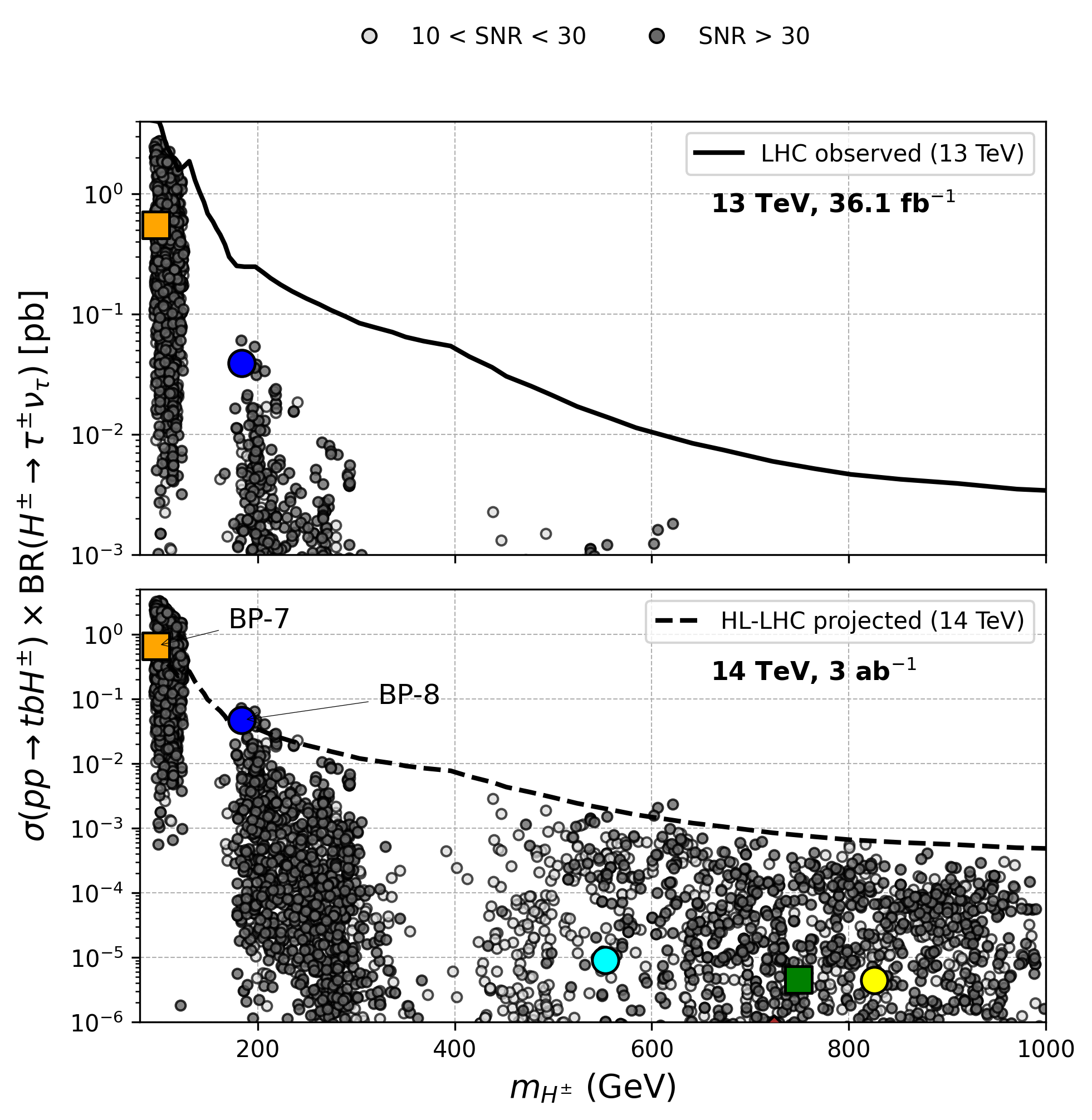}
    \caption{Same as Fig.~\ref{fig:Collider_HZZ_Hhh}, but for $tbH^\pm$ associated production rates times decay BRs in the $H^\pm\to \tau^\pm\nu_\tau$ channel. The gaps in the distribution of points are related to having plotted only configurations with SNR $>10$ (see Appendix~\ref{sec:correlation}).}
    \label{fig:Collider_tbHc}
\end{figure}

\subsection{Collider Sensitivity}
\label{subsec:collider_interpretation}

In this subsection, we discuss the structure and qualitative features of the collider results presented in Figs.~\ref{fig:Collider_HZZ_Hhh}--\ref{fig:Collider_tbHc}, focusing on the production and decay channels most relevant for the A2HDM. The underlying parameter space has been classified into eight distinct datasets, as introduced in Fig.~\ref{fig:1st_order_frac}, all of which satisfy the full set of theoretical requirements and current experimental constraints.

To assess the collider complementarity of the scenarios of interest, we restrict our attention to the parameter points that exhibit a strong GW signal with a value $\mathrm{SNR} > 10$. These points constitute a phenomenologically well-motivated subset of the full scan and form the basis of the collider projections shown in this section.

The figures are presented as plots of the production cross sections $\sigma\times$BR  versus the corresponding  non-SM (pseudo)scalar mass, enabling a direct comparison between theoretical predictions and experimental limits. For clarity, the parameter points from all eight datasets are combined in each plot, highlighting the overall collider-relevant structure of the A2HDM parameter space rather than its dataset-specific features.  
Theoretical predictions of such production and decay rates at 13 and 14~TeV obtained from the parameter space scan are shown as gray points, corresponding to parameter sets that satisfy all theoretical and experimental constraints discussed in previous sections, wherein the light gray colour corresponds to 10 $<$ SNR $<$ 30 and the dark gray one to SNR $>30$.
The solid black curve represents the current upper limit derived from LHC searches at $\sqrt{s} = 13~\mathrm{TeV}$, as reported by the ATLAS and CMS collaborations for the corresponding final state. These limits are shown at the luminosity of the existing dataset and provide a reference for the present collider sensitivity. The dashed black curve indicates the projected exclusion reach at the HL-LHC, obtained by extrapolating the    $\sqrt{s} =13$~TeV limits to 14~TeV and an integrated luminosity of $\mathcal{L} = 3~\mathrm{ab}^{-1}$ using the scaling procedure previously described. In short, the purpose of this exercise is to find points between the solid and dashed curves (i.e., configurations of the A2HDM that are not presently excluded and can be discovered in the future) with SNR $>10$, so that one can claim that the A2HDM can simultaneously produce detectable signals in terms of both GWs and collider events.

For the neutral Higgs channels produced via ggF, $gg \to  H, A$, the scaling of the experimental limits reflects the enhancement of the gluon-gluon parton luminosity at higher collider energy, which becomes increasingly important for larger Higgs masses. As a result, the separation between the current LHC limit and the HL-LHC projection generally grows with increasing $m_{H, A}$, illustrating the improved sensitivity of the HL-LHC to particularly heavy (pseudo)scalar states, as seen in Figs.~\ref{fig:Collider_HZZ_Hhh}--\ref{fig:Collider_Htautau_AZh}.
In the case of charged Higgs production, $pp \to tbH^\pm$, followed by the decay $H^\pm \to \tau^\pm \nu_\tau$, a substantial accessible mass range, both below $m_t$ and above it, at the HL-LHC is clearly visible in Fig.~\ref{fig:Collider_tbHc}.

Examining the collider results channel by channel, as shown in Figs.~\ref{fig:Collider_HZZ_Hhh}--\ref{fig:Collider_tbHc}, we observe a common qualitative pattern across all heavy Higgs searches. In the top panels, corresponding to the current Run 3 sensitivity at $\sqrt{s}=13~\mathrm{TeV}$, all parameter points lie well below the experimental exclusion limits, indicating that present data do not really constrain these scenarios\footnote{With the exception of one point in forthcoming Fig.~\ref{fig:Collider_HZZ_Hhh} (right plot), which we will eventually elect to be one of our BPs (BP-5), which could actually be discovered already at Run 3 of the LHC.}. In contrast, the bottom panels demonstrate that a substantial fraction of the same parameter space becomes accessible at the HL-LHC, with many points lying well above the projected sensitivity curves\footnote{Notice that the points in the lower and upper frames are the same, yet, their distribution is different, owing to the different scaling factors for each point (as previously explained).}.
In a few of the previously discussed channels, i.e., $gg\to H \to W^+W^-$ and $pp \to t \bar{t} H, A$ followed by $H, A \to t \bar{t}$, there is no sensitivity even at the HL-LHC, though, since all the considered parameter points remain below the projected reach: therefore, we do not include the corresponding results in the remainder of our paper. In contrast, the $gg\to H \to ZZ$ (left panel of Fig.~\ref{fig:Collider_HZZ_Hhh}) and $gg\to H \to hh$ (right panel of Fig.~\ref{fig:Collider_HZZ_Hhh}) production and decay modes provide a more promising probe, allowing sensitivity to heavy scalar masses above $m_H \gtrsim 500~\mathrm{GeV}$, with additional isolated points potentially accessible in the intermediate mass region $300~\mathrm{GeV} \lesssim m_H \lesssim 500~\mathrm{GeV}$.
The $\tau^+\tau^-$ final state, shown in the left panel of Fig.~\ref{fig:Collider_Htautau_AZh}, predominantly probes the lower mass region, with sensitivity extending to the interval $90~\mathrm{GeV} \lesssim m_H \lesssim 150~\mathrm{GeV}$. Moreover, $pp\to H \to \tau^+\tau^-$ also exhibits enhanced sensitivity at the HL-LHC for heavier scalars with $m_H \gtrsim 400~\mathrm{GeV}$. In the right panel of Fig.~\ref{fig:Collider_Htautau_AZh}, one can see the results for the ggF production of CP-odd scalar $A$; the $gg \to A \to Zh$ channel consistently offers strong discovery prospects at the HL-LHC, particularly for masses $m_A \gtrsim 400~\mathrm{GeV}$. This channel is among the most effective probes of the extended Higgs sector in the considered parameter space of the A2HDM.
Furthermore, charged Higgs searches via associated production, $pp \to tbH^\pm$ followed by the decay $H^\pm \to \tau^\pm \nu_\tau$ (Fig.~\ref{fig:Collider_tbHc}), display sensitivity over a wide mass range. In particular, a large fraction of the parameter space with $m_{H^\pm} \gtrsim 90~\mathrm{GeV}$ can be probed at the HL-LHC, highlighting the strong synergy between charged and neutral Higgs searches.

To enable further phenomenological work, we finish this section by introducing eight BPs. We first focus on the $H \to ZZ$ decay mode (left panel of Fig.~\ref{fig:Collider_HZZ_Hhh}) and select points with $10 < \text{SNR} < 30$, which are likely to be accessible at the HL-LHC via this channel. We then highlight these BPs across other possible decay modes (see Figs.~\ref{fig:Collider_HZZ_Hhh}–\ref{fig:Collider_tbHc}) to assess their potential for detection through multiple searches. This selects BP-1 through BP-6,
wherein the additional Higgs masses are rather heavy.
Therefore, to access low-mass (pseudo)scalars, we choose two additional configurations (BP-7 and BP-8), the first one can be accessed through both $gg\to H \to \tau^+ \tau^-$ (the left panel of Fig.~\ref{fig:Collider_Htautau_AZh}) and $pp\to t b H^{\pm} \to \tau \nu_{\tau}$ (Fig.~\ref{fig:Collider_tbHc}) while the second  is only through the latter. Hence, while BP-1 to BP-6 and BP-8
correspond to parameter space configurations of the A2HDM that are accessible through more than one  discovery channels at the HL-LHC, 
BP-7 is really only accessible through one channel, but we have included it in our selection as it gives access to the region of A2HDM parameter space realising the so-called inverted mass hierarchy in the neutral CP-even Higgs sector, i.e., $m_H<125.2$ GeV. (Also, do recall that BP-5 may offer some sensitivity already by the end of Run 3 of the LHC.)
The selected BPs are summarised in Tab.~\ref{tab:bp_summary}, which symbols are highlighted in the discussed figures (when visible) and also labelled (when discoverable). 

\begin{table}[h!]
\centering
\hspace*{-10mm}\scalebox{1.0}{
\begin{tabular}{cccccccccc}
\toprule
BP & $m_{H}$ & $m_{A}$ & $m_{H^\pm}$ & SNR & $H \to ZZ$ & $H \to \tau^+ \tau^-$ & $H \to hh$ & $A \to Zh$ & $H^{\pm} \to \tau \nu_{\tau}$ \\
\midrule
\bpone\ BP-1   & 833.6 & 772.1 & 826.0 & 27.4 & \checkmark & $\times$ & \checkmark & \checkmark & $\times$ \\
\bptwo\ BP-2   & 741.7 & 703.6 & 748.8 & 13.3 & \checkmark & $\times$ & \checkmark & \checkmark & $\times$ \\
\bpthree\ BP-3 & 705.6 & 672.7 & 723.9 & 28.8 & \checkmark & $\times$ & \checkmark & \checkmark & $\times$ \\
\bpfour\ BP-4  & 624.1 & 723.7 & 560.9 & 46.3 & \checkmark & $\times$ & \checkmark & $\times$ & $\times$ \\
\bpfive\ BP-5  & 330.8 & 584.4 & 593.2 & 27.3 & \checkmark & $\times$ & \checkmark & $\times$ & $\times$ \\
\bpsix\ BP-6   & 564.9 & 638.3 & 552.9 & 14.8 & \checkmark & $\times$ & $\times$ & \checkmark & $\times$ \\
\bpseven\ BP-7 & 95.5  & 215.4 & 96.9  & 27.3 & $\times$ & $\times$ & $\times$ & $\times$ & \checkmark \\
\bpeight\ BP-8 & 122.8 & 91.1  & 183.8 & 12.4 & $\times$ & \checkmark & $\times$ & $\times$ & \checkmark \\
\bottomrule
\end{tabular}}
\caption{Summary of BPs with Higgs masses and decay channels offering the possibility of being probed at the HL-LHC. 
In particular, BP-1, BP-2, and BP-3 can be explored through $gg\to H \to ZZ$, $gg\to H \to hh$, and $gg\to A \to Zh$. In contrast, BP-4 and BP-5 can be probed via $gg\to H \to ZZ$ and $gg\to H \to hh$ while BP-6 shows potential for detection through $gg\to H \to ZZ$ and $gg\to A \to Zh$. Finally, BP-7 and BP-8 can both be probed through $pp\to tb H^{\pm} \to \tau \nu_{\tau}$ with BP-8 accessible also via $gg\to H \to \tau^+ \tau^-$.}
\label{tab:bp_summary}
\end{table}

The collider analysis presented in the previous subsection highlights the interplay between
GW observations and searches for extended Higgs sectors at the LHC.

From the collider perspective, heavy Higgs searches are most sensitive to regions of
parameter space where the ggF production cross-section and the branching ratios
into EW gauge bosons or Higgs pairs are sufficiently enhanced.
As shown in our results, only a small fraction (typically at the level of a few percent)
of the parameter points are compatible with a strong FOEWPT and yields $\sigma\times \mathrm{BR}$
values within the projected reach of the HL-LHC.

Interestingly, there exist regions of the parameter space where collider sensitivities are
limited due to suppressed production and/or decay rates or challenging signatures but are accessible to future space-based
GW experiments.
This makes GW observations a powerful and complementary probe of the
A2HDM.

The combined use of collider searches and GW measurements, therefore, allows
for a more comprehensive exploration of the A2HDM parameter space than either approach
alone.
While collider experiments directly test the particle spectrum and interactions of the
model, GWs provide a unique window into the dynamics of EWSB in the early universe.
Together, these probes offer a consistent and mutually reinforcing strategy to test the
viability of extended Higgs sectors and their cosmological implications, chiefly of the A2HDM
exploited here.

\section{Summary and Conclusions}
\label{sec:conclusion}

The SM fails to trigger a strong FOEWPT. With the Higgs mass fixed at about 125 GeV, it predicts a smooth cross-over rather than a genuine first-order transition. This makes the SM unable to generate the strong departure from equilibrium needed for the production of observable GW signals from the EWPT. In turn, additional fields and/or interactions are needed to modify the thermal history of the Higgs sector and make a strong FOEWPT possible.

\bl{Extended (pseudo)scalar sectors offer a minimal way to achieve this. In fact, additional Higgs degrees of freedom can significantly modify the finite-temperature dynamics of EWSB while preserving the renormalisability of the underlying Lagrangian. The 2HDM is one of the simplest such extensions, providing enough extra (pseudo)scalar states that can drive a strong FOEWPT while remaining compatible with current theoretical and experimental constraints.}

Within the 2HDM, the aligned version (A2HDM) is especially appealing with a rich Yukawa structure, while avoiding tree-level FCNCs without imposing discrete symmetries. At the same time, the additional Higgs states remain free to modify the scalar potential and support a strong FOEWPT. This combination of theoretical economy and phenomenological viability thus makes the A2HDM a compelling framework in which to study the generation of detectable GWs from the EWPT and their interplay with collider signatures.

In this BSM setting, we have explored the parameter space, which we have unfolded into eight regions, each characterised by a particular Higgs mass spectrum, able to trigger a strong FOEWPT with the presence of GW signals that can be accessible by LISA (and potentially other future interferometer experiments). In this context, the parameter region in which all BSM Higgs states are heavier than the SM Higgs emerges as the most favourable scenario,  since this configuration yields the largest fraction of parameter points with high SNRs at LISA, thereby producing detectable GWs with very low peak frequencies. In contrast, the scenario where all BSM Higgs states are lighter than the SM Higgs is the least promising and can only marginally be probed by LISA. 

Simultaneously, over the same parameter space where such GW signals are realised, we have highlighted which Higgs production and decay processes, unconstrained by Run 3 data, can be accessed at the HL-LHC, thereby setting the stage for a complementary pursuit of the A2HDM from 2030/35 onwards. Remarkably, we have found that the A2HDM is particularly viable at the CERN collider in the parameter space regions where all additional (pseudo)scalar Higgs states are heavier than the SM-like Higgs boson, although scenarios with light charged and/or neutral Higgs states also offer some sensitivity. For these mass spectra, we have studied the processes that can be used at the HL-LHC to probe such a BSM scenario. Focusing on the most sensitive production and decay channels of non-SM Higgs states, we have compared predicted signal rates with current Run 3 LHC limits as well as HL-LHC projections. While much of the parameter space remains unconstrained presently, a significant fraction of it can be probed in the future through $gg\to H \to ZZ$, $gg\to H \to hh$, $gg\to H \to \tau^+ \tau^-$, $gg\to A \to Zh$ and $pp \to tbH^\pm$ with $H^\pm \to \tau^\pm \nu_\tau$. Highlighting eight BPs with $\text{SNR} > 10$, for both heavy and light additional (pseudo)scalar states, both neutral and charged ones, we have found that one or more of these can be accessed at the HL-LHC, thus providing the advocated complementary probes of the A2HDM parameter space favoured by a strong FOEWPT and GW signals.

A further motivation for studying such a scenario is offered by the matter-antimatter asymmetry of the Universe. Extended (pseudo)scalar sectors can in principle also introduce new CP-violating phases, both spontaneous and explicit, while supporting a strong FOEWPT. As such, the A2HDM also provides a compelling framework for embedding EW baryogenesis which we are currently implementing.

\section*{Acknowledgments}
AD is supported by the Carl Trygger Foundation under the project CTS 23:2930. The work of AK, LDR and SDC has been supported by the research grant number 20227S3M3B “Bubble Dynamics in Cosmological Phase Transitions” under the program PRIN 2022 of the Italian Ministero dell’Università e Ricerca (MUR). AK also thanks LHCPheno group at IFIC(UV-CSIC), Valencia, Spain, for providing the computing facilities.
SM is supported in part through the NExT Institute and STFC Consolidated Grant ST/X000583/1. 
SM thanks Andr\'e Pousette and Carlo Tasillo for innumerable enlightening discussions on the physics of EWPTs and GWs. LDR thanks Ville Vaskonen for useful discussions. MR thanks the Theoretical Particle Physics group at Uppsala University for hospitality during completion of parts of this work.
SDC would like to thank the Galileo Galileo Institute (GGI) for Theoretical Physics for the kind hospitality. AC thanks the CFTC – Center for Theoretical and Computational Physics at Lisbon University for their hospitality during part of this work.

\appendix

\section{BPL vs DBPL Comparison}
\label{sec:BPL}

The GW spectrum can be  estimated also by implementing a Broken Power Law (BPL). In this method, the GW spectrum reads~\cite{Caprini:2024hue}:
\begin{equation}
    h^2 \Omega_{\rm GW}^{\rm BPL}= h^2 \Omega_{p} \left(\frac{f}{f_p}\right)^{\tilde n_1}\left[\frac{\tilde n_1-\tilde n_2}{\tilde n_1 \left(\frac{f}{f_p}\right)^{\tilde a_1}-\tilde n_2}\right]^{\frac{\tilde n_1-\tilde n_2}{\tilde a1}},
\end{equation}
where $f_p$ and $\Omega_p$ are respectively the frequency and amplitude corresponding to the peak of the spectrum. Using the values of $\tilde n_1=3$, $\tilde n_2=-4$ and $\tilde a_1=2$, one achieves the formula commonly used for sound waves in the literature~\cite{Carena:2025flp} as:
\begin{equation}
    h^2\Omega_{\rm sw}(f)=h^2\Omega_{\rm sw}^{\rm peak } \left(\frac{f}{f_{\rm sw}}\right)^3 \left[\frac{7}{4+3\left(\frac{f}{f_{sw}}\right)^2}\right]^{7/2}.
\end{equation}
The peak frequency and peak amplitude of the sound wave observed today read as follows:
\begin{eqnarray}
    f_{\rm sw}
= 1.9 \times 10^{-5}\,{\rm Hz}
\left( \frac{1}{v_w} \right)
\left( \frac{\beta}{H} \right)
\left( \frac{g_*}{100} \right)^{1/6}
\left( \frac{T_n}{100~{\rm GeV}} \right),\\
    h^2 \Omega_{\rm sw}^{\text{peak}}
= 2.65 \times 10^{-6}
\left( \frac{\kappa_{\rm sw}\,\alpha}{1 + \alpha} \right)^2
\left( \frac{H}{\beta} \right)
\left( \frac{100}{g_*} \right)^{1/3}
(H \tau_{\rm sw})\, v_w,
\end{eqnarray}
with $\kappa_{\text{SW}}$ defined in Eq.~\eqref{eq:kcoeff}.
The dimensionless factor $H \tau_{\text{sw}}$ accounts for the effective duration of the sound wave source relative to the Hubble time, i.e.,
\begin{equation}
    H \tau_{\rm sw}
= \min \left[
1,\,
(8\pi)^{1/3}
\left( \frac{\max(v_w, c_s)}{\beta/H} \right)
\left( \frac{4}{3}
\frac{1 + \alpha}{\kappa_{\rm sw}\,\alpha}
\right)^{1/2}
\right].
\end{equation}

\begin{figure}[t!]
    \centering
    \includegraphics[width=0.7\linewidth]{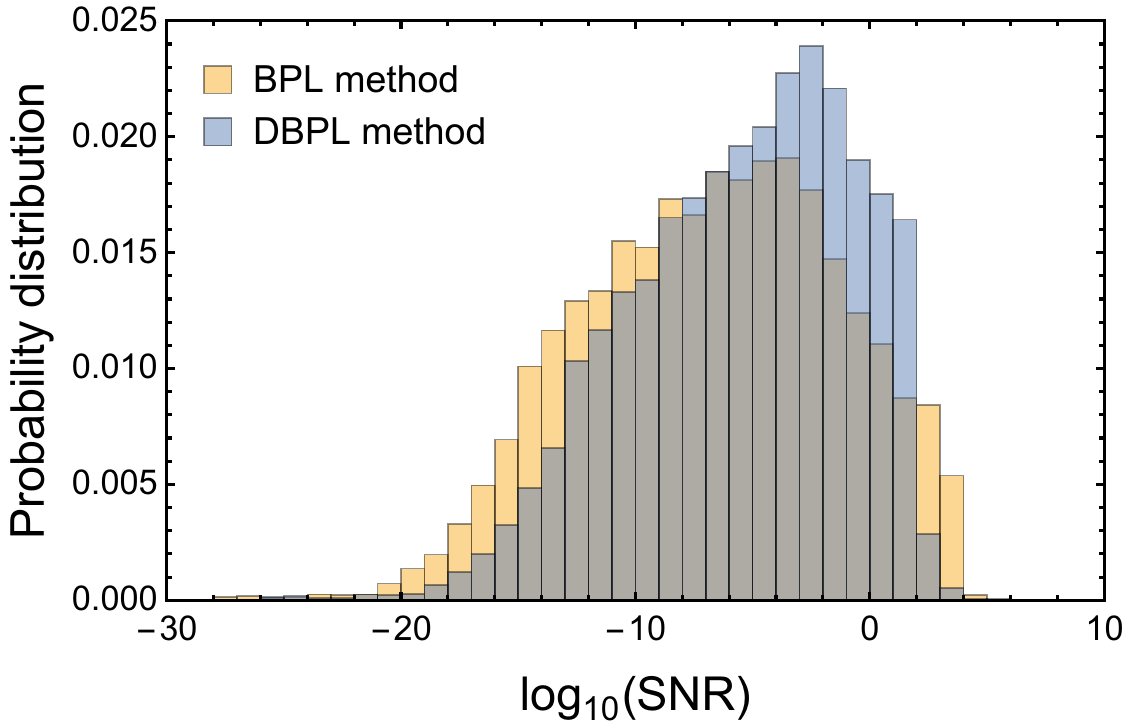}
    \caption{Probability distribution of $\log_{10}(\rm SNR)$ obtained by using the BPL method (yellow bars) and the DBPL method (blue bars) for the ``All Heavy'' scenario.}
    \label{fig:snr_dist}
\end{figure}

\begin{figure}[h!]
\centering
\includegraphics[width=0.4\linewidth]{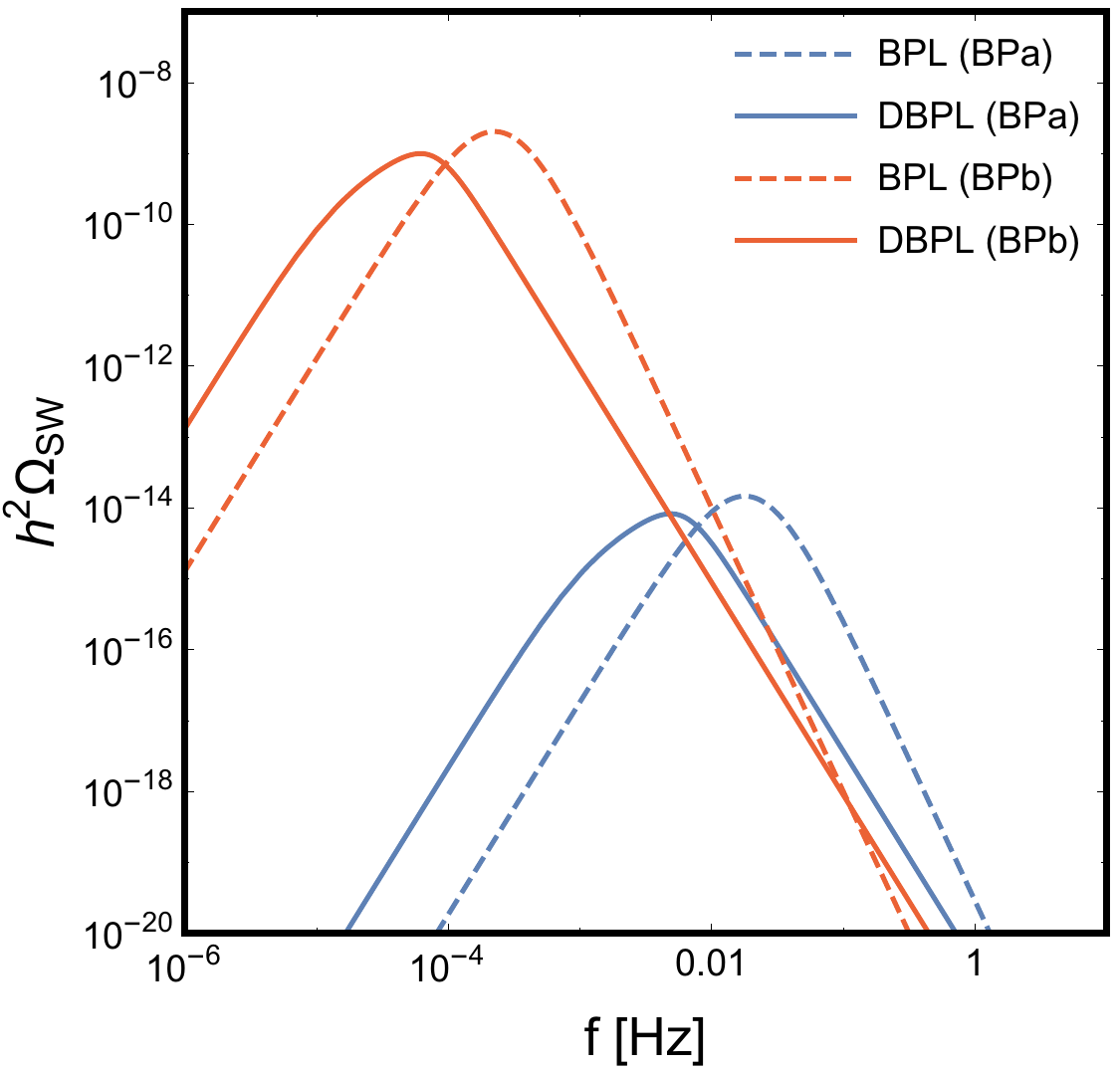}
\hfil
\includegraphics[width=0.4\linewidth]{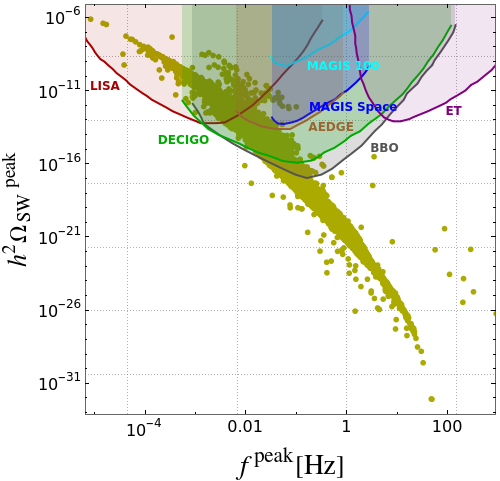}
\caption{Left panel: GW spectra for two BPs (BPa and BPb defined in the text). The dashed curves indicate the spectra obtained with the  BPL method while the solid curves represent the same in the  DBPL method. Right Panel: GW peak amplitude versus peak frequency obtained with the BPL method for the ``All Heavy'' scenario.}
\label{fig:GWspectra_BPL}
\end{figure}

In Fig.~\ref{fig:snr_dist}, we present the distribution of $\log_{10}(\mathrm{SNR})$ obtained using both methods in the scenario with all (pseudo)scalars states heavier than the SM-like Higgs boson. The yellow histogram corresponds to the distribution obtained  using the   BPL approach while the blue histogram shows the results from the DBPL method. The distributions of $\log_{10}(\rm SNR)$ in both methods are quite similar, although the values of SNR for individual parameter space points are different.

In Fig.~\ref{fig:GWspectra_BPL}, left panel, we present the comparison of GW spectra in both the methods considering the ``All Heavy'' scenario. In the left panel we plot those for two particular BPs: BPa (blue) and BPb (red). {To BPa corresponds a relatively small SNR ($\sim 0.32$), whereas BPb represents a point with a significantly larger SNR ($\sim 50.90$). The difference in the peak amplitudes of the GW spectra for these two BPs reflects the corresponding difference in their SNR values.}
 Notice that the estimated peak frequency in DBPL is smaller compared to the same in BPL. In the right panel of the same figure, we plot the peak amplitude versus the peak frequency  for all the allowed points in ``All Heavy'' scenario using the BPL method. Comparing this with Fig.~\ref{fig:GWspectra}(a), one can notice that the cluster of points has indeed shifted rightwards if  the BPL method is used. {A similar trend has been noticed for the other mass spectra too.}

{{In short, in the scenario most effectively enabling a strong FOEWPT (``All Heavy'') in the
A2HDM, the differences between two mainstream methods to compute SNRs for the ensuing GWs are statistically somewhat marginal, so as to corroborate the solidity of our approach to cosmological dynamics, however, noticeable differences persist in the peak frequency distribution.}}

\section{The EW Baryogenesis Criterion: $v_c/T_c$}
\label{sec:vcbytc}

The connection between the EWPT strength and EW baryogenesis is encapsulated by the ratio $v_c/T_c$. In particular, the baryon asymmetry is generated by expanding bubble walls during a strong FOEWPT, where out-of-equilibrium conditions are satisfied ~\cite{Cohen:1993nk, Morrissey:2012db, Trodden:1998ym, vandeVis:2025efm}. CP-violating interactions at the wall produce chiral charge densities, which are partially converted into a baryon number by EW sphalerons in the symmetric phase. In the following, we assess the compatibility of the A2HDM with EW baryogenesis by considering the sphaleron wash-out condition~\cite{Kuzmin:1985mm, Shaposhnikov:1986jp, Shaposhnikov:1987pf, Shaposhnikov:1987tw}. A preservation of the baryon asymmetry requires a strong FOEWPT, conventionally expressed as $v_c/T_c \gtrsim 1$, which ensures the suppression of baryon-number violating sphaleron processes in the broken phase.
\vspace*{-1.5mm}
\begin{figure}[h!]
\centering
\includegraphics[trim={0cm 0.8cm 6cm 6cm },clip,scale=1.0]{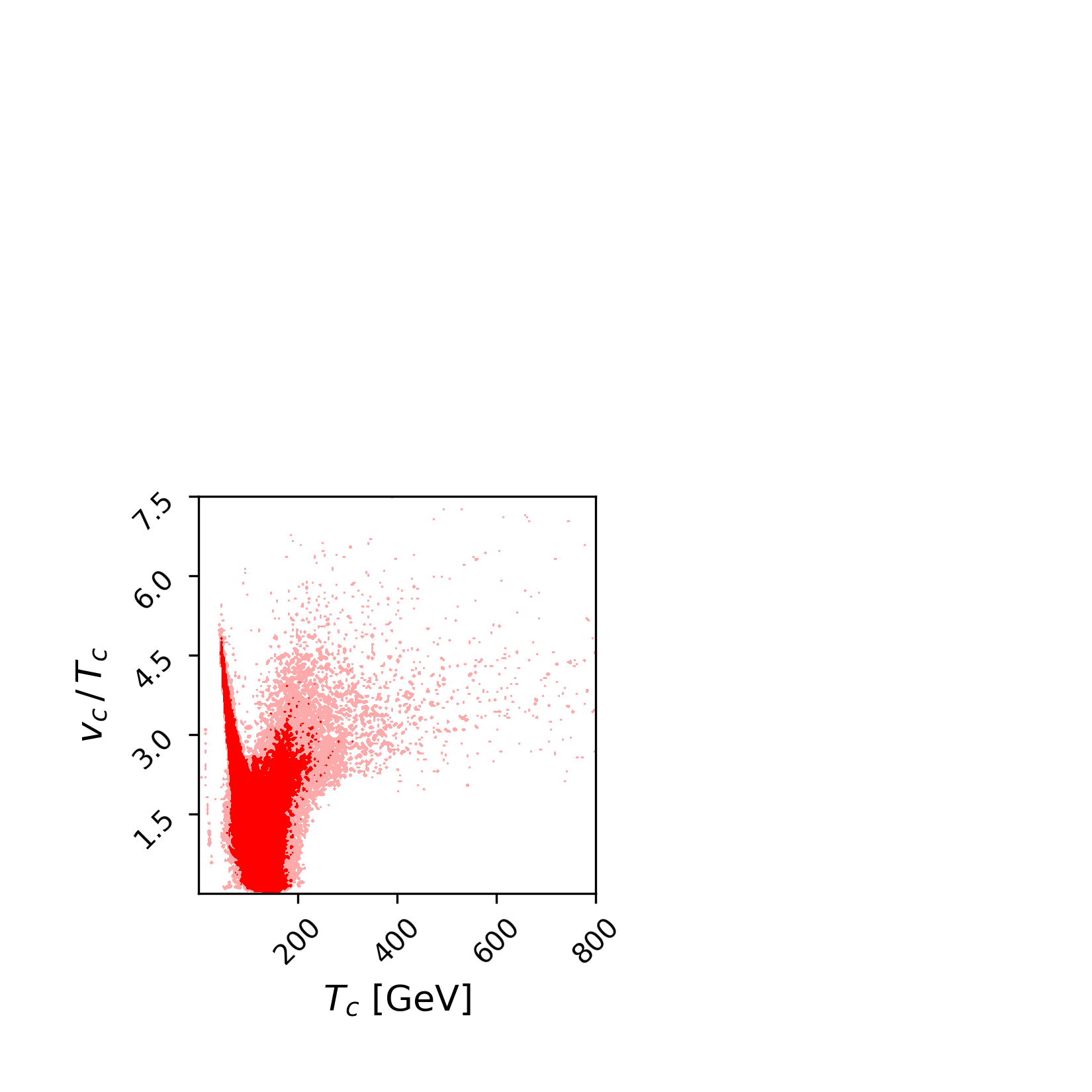}
\vspace*{-1.5mm}
\caption{Scatter plots of $v_c/T_c$ vs $T_c$ combining all the regions of A2HDM parameter space. Dark red indicates 95\% probability region while light red signify 99.7\% probability region.}
\label{fig:vcTc_Tc_params}
\end{figure}

\vspace*{-1mm}
In Fig.~\ref{fig:vcTc_Tc_params}, we present the ratio $v_c/T_c$ as a function of $T_c$, obtained by combining the parameter space regions from all eight scenarios. The condition $v_c/T_c \gtrsim 1$, required for a strong FOEWPT, is satisfied in each of the scenarios considered. In the figure, the dark red region corresponds to the $95\%$ probability interval, while the light red region represents the $99.7\%$ probability interval. Although $T_c$ spans a relatively wide range, the $95\%$ probability region predominantly satisfies $T_c \lesssim 220$ GeV. A substantial portion of the parameter space yields values of $v_c/T_c$ exceeding the unity threshold, with the $95\%$ probability region reaching values as large as $v_c/T_c \sim 5$. Furthermore, even for $T_c$ values exceeding 200 GeV, the strong FOEWPT condition remains robustly satisfied within the $99.7\%$ probability region. 

{This result indicates that a substantial region of the A2HDM parameter space can satisfy the EW baryogenesis criterion by generating a strong FOEWPT. With the inclusion of CP-violating parameters in both the scalar potential and the Yukawa interactions, the A2HDM will emerge as a viable framework for explaining the observed baryon asymmetry of the Universe via EW baryogenesis.}

\vspace*{-1mm}
\section{SNR vs masses and Lagrangian parameters}
\label{sec:correlation}
\vspace*{-2mm}

\begin{figure}[h!]
    \centering
   \includegraphics[trim={0cm 0cm 6.0cm 6.0cm},clip,width=\linewidth]{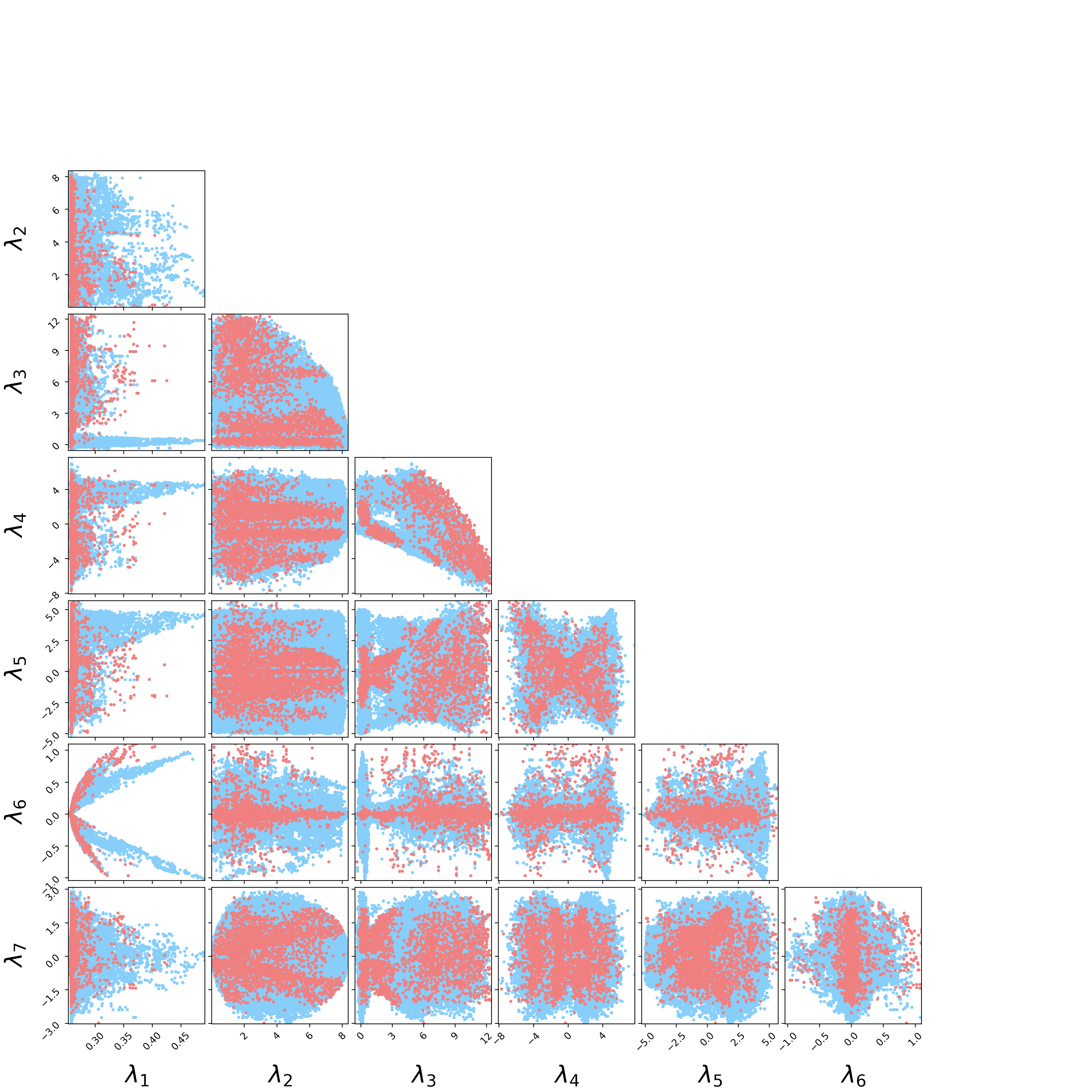}
 \caption{Correlations among quartic couplings for all eight scenarios combined. The blue colour represents parameter points with SNR smaller than 10 (in DBPL) while the red ones indicate the reverse.}
    \label{fig:lam_snr_gr10}
\end{figure} 

\begin{figure}[h!]
    \centering
   \includegraphics[trim={0cm 0cm 6.0cm 6.0cm},clip,width=0.45\linewidth]{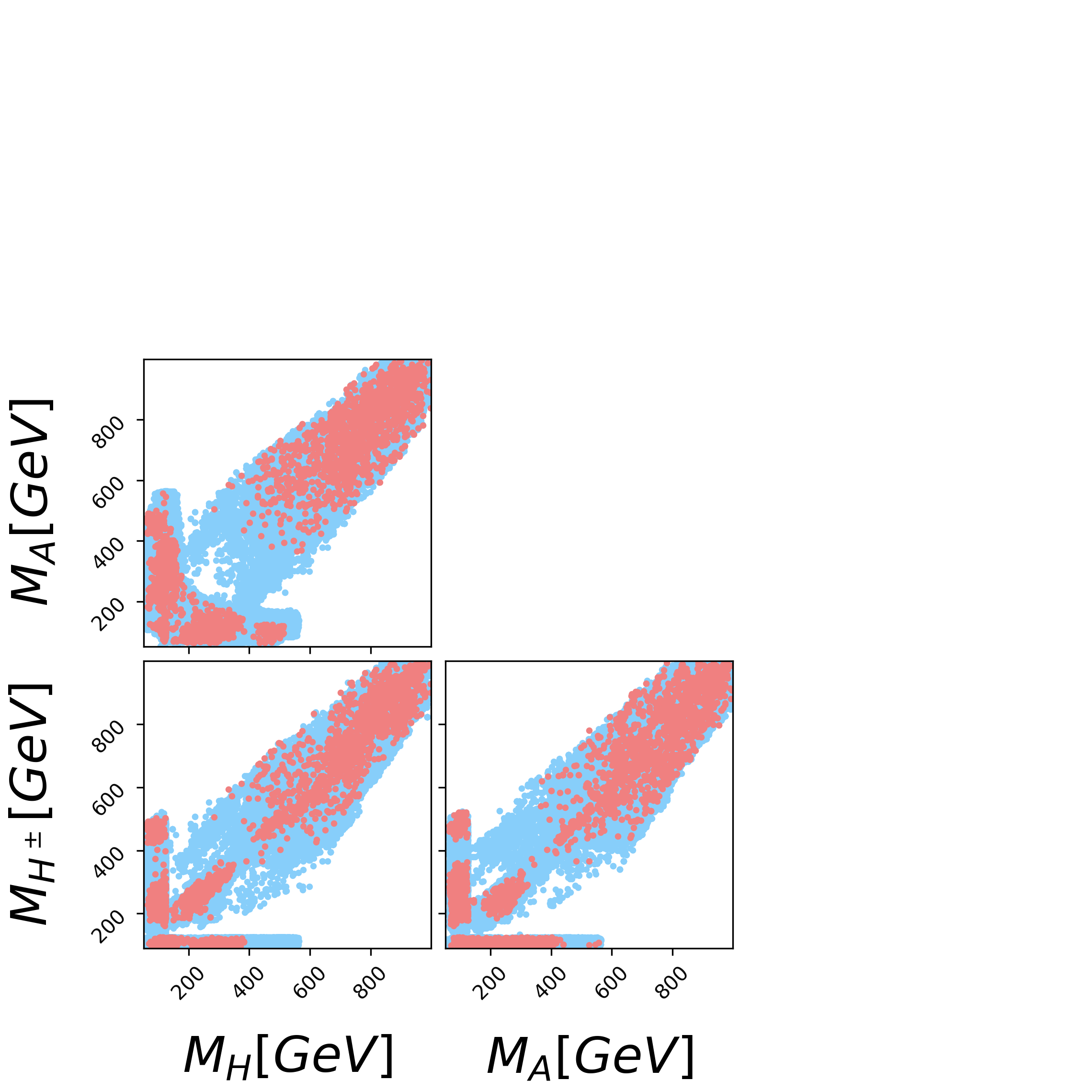}
   \hfil
   \includegraphics[trim={0cm 0cm 6.0cm 6.0cm},clip,width=0.45\linewidth]{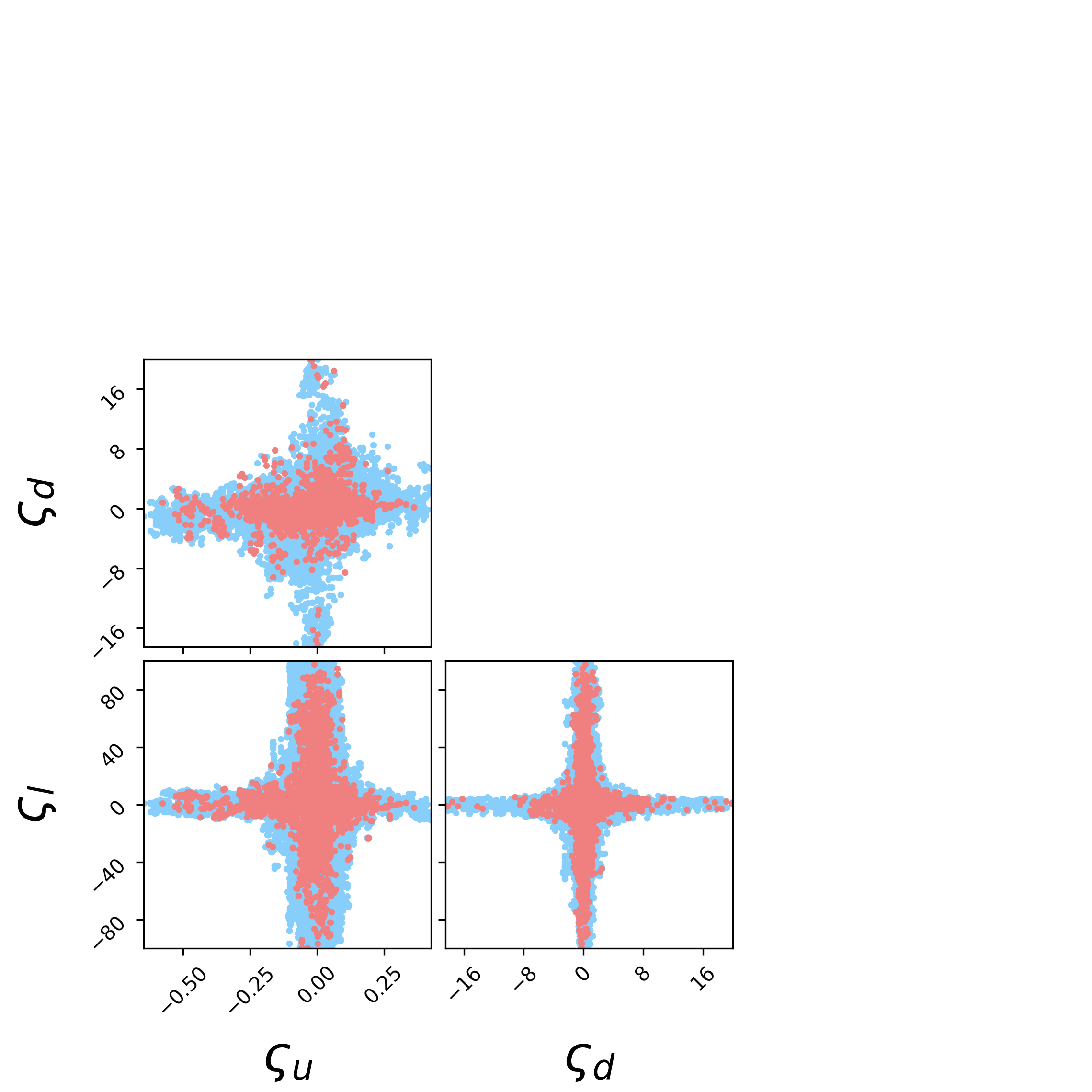}
    \caption{Correlations among the masses of the BSM (pseudo)scalars (left) and alignment parameters (right) for all eight scenarios combined. The blue colour represents parameter points with SNR smaller than 10 (in DBPL) while the red ones indicate the reverse.}
    \label{fig:M_sig_snr_gr10}
\end{figure}

In Figs.~\ref{fig:lam_snr_gr10} and \ref{fig:M_sig_snr_gr10}, we present the correlations among quartic couplings, masses, alignment parameters and quartic couplings of the allowed points combining all eight parameter space scenarios of the A2HDM, which are in agreement with Refs.~\cite{Karan:2023kyj,Coutinho:2024zyp,Karan:2024kgr,Karan:2023xze,Miralles:2025kes,Coutinho:2024vzm}. While the blue colour represents parameter points with SNR less than 10 (in DBPL), the red ones indicate the opposite. A notable feature of Fig.~\ref{fig:M_sig_snr_gr10} is the fact that the request of having SNR $>10$ introduces gaps in the mass distributions of the $H,A$ and $H^\pm$ states, as reflected in Figs.~\ref{fig:Collider_HZZ_Hhh}--\ref{fig:Collider_tbHc}.

\FloatBarrier

\bibliographystyle{JHEP}
\bibliography{biblio}{}
\end{document}